%% file: paper.tex
\documentclass{aa}
\usepackage{graphicx}
\usepackage{bm}
\usepackage{times}
\usepackage{amsmath}
\usepackage{amssymb}
\usepackage{natbib}
\usepackage{color}
\usepackage{mathrsfs}
\usepackage[colorlinks,allcolors=blue]{hyperref}
\usepackage{url}
\usepackage{multirow}
\bibpunct{(}{)}{;}{a}{}{,}
\graphicspath{{./fig/}{./png/}}
\input macros

\begin{document}

\authorrunning{K\"apyl\"a}
\titlerunning{Prandtl number dependence of compressible convection}

   \title{Prandtl number dependence of compressible convection:\\ Flow statistics and convective energy transport}

   \author{P. J. K\"apyl\"a
          \inst{1,2}
          }

   \institute{Georg-August-Universit\"at G\"ottingen, Institut f\"ur 
              Astrophysik, Friedrich-Hund-Platz 1, D-37077 G\"ottingen,
              Germany
              email: \href{mailto:pkaepyl@uni-goettingen.de}{pkaepyl@uni-goettingen.de}
   \and
              Nordita, KTH Royal Institute of Technology and Stockholm
              University, Stockholm, Sweden
}

\date{\today}

\abstract{The ratio of kinematic viscosity to thermal diffusivity, the
  Prandtl number, is much smaller than unity in stellar convection zones.
   }%
   {The main goal is to study the statistics of convective flows and
     energy transport as functions of the Prandtl number.
   }%
   {Three-dimensional numerical simulations of compressible
     non-rotating hydrodynamic convection in Cartesian geometry are
     used. The convection zone (CZ) is embedded between two stably
     stratified layers. The dominant contribution to the diffusion of
     entropy fluctuations comes in most cases from a subgrid scale
     (SGS) diffusivity whereas the mean radiative energy flux is
     mediated by a diffusive flux employing Kramers opacity
     law. Statistics and transport properties of up- and downflows are
     studied separately.
   }%
   {The volume-averaged rms velocity increases with decreasing Prandtl
     number. At the same time the filling factor of downflows
     decreases and leads to, on average, stronger downflows at lower
     Prandtl numbers. This results in a strong dependence of
     convective overshooting on the Prandtl number. Velocity power
     spectra do not show marked changes as a function of Prandtl
     number except near the base of the convective layer where the
     dominance of vertical flows is more pronounced. At the highest
     Reynolds numbers the velocity power spectra are more compatible
     with the Bolgiano-Obukhov $k^{-11/5}$ than the Kolmogorov-Obukhov
     $k^{-5/3}$ scaling. The horizontally averaged convected energy
     flux ($\mFconv$), which is the sum of the enthalpy ($\mFenth$)
     and kinetic energy fluxes ($\mFkin$), is independent of the
     Prandtl number within the CZ. However, the absolute values of
     $\mFenth$ and $\mFkin$ increase monotonically with decreasing
     Prandtl number. Furthermore, $\mFenth$ and $\mFkin$ have opposite
     signs for downflows and their sum $\mFconv^{\downarrow}$
     diminishes with Prandtl number. Thus the upflows (downflows) are
     the dominant contribution to the convected flux at low (high)
     Prandtl number. These results are similar to those from
     Rayleigh-Ben\'ard convection in the low Prandtl number regime
     where convection is vigorously turbulent but inefficient at
     transporting energy.
   }%
   {The current results indicate a strong dependence of convective
     overshooting and energy flux on the Prandtl number. Numerical
     simulations of astrophysical convection often use Prandtl number
     of unity because it is numerically convenient. The current
     results suggest that this can lead to misleading results and that
     the astrophysically relevant low Prandtl number regime is
     qualitatively different from the parameters regimes explored in
     typical contemporary simulations.
   }%

  \maketitle


\section{Introduction}

The flows in solar and stellar CZs are characterised by very high
Reynolds and P\'eclet numbers $\Rey=u\ell/\nu$ and $\Pe=u\ell/\chi$,
respectively, where $u$ and $\ell$ are typical velocity and length
scales, and $\nu$ and $\chi$ are the kinematic viscosity and thermal
diffusivity \citep[e.g.][]{O03,K11,2020RvMP...92d1001S}. This implies
very vigorous turbulence which means that resolving all scales down to
the Kolmogorov scale is infeasible \citep[e.g.][]{CS86}. Moreover, the
ratio $\Pe/\Rey$, which is the Prandtl number $\Pr=\nu/\chi$, is
typically much smaller than unity in stellar CZs
\citep[e.g.][]{2019ApJ...876...83A}. For example, values of the order
of $\Pra\lesssim10^{-6}$ are typical in the solar CZ. Most numerical
simulations, however, are made with Prandtl number of the order of
unity because greatly differing viscosity and thermal diffusivity
would lead to a wide gap in the smallest physically relevant scales of
velocity and temperature. Hence, reaching high $\Rey$ and $\Pe$
simultaneously in simulations with low $\Pra$ is prohibitively
expensive \citep[e.g.][]{2017LRCA....3....1K}.

Recently it has become clear that current simulations do not capture
some basic features of solar convection accurately. Comparisons of
helioseismic and numerical studies suggest that the simulations
produce significantly higher velocity amplitudes at large horizontal
scales \citep[e.g.][]{HDS12,2016AnRFM..48..191H}. While discrepancies
exist also between helioseismic methods \citep[see, e.g.][]{GHFT15},
there is another, more direct piece of evidence from simulations:
global and semi-global simulations with solar luminosity and rotation
rate preferentially lead to anti-solar differential rotation with a
slower equator and faster poles
\citep[e.g.][]{FF14,KKB14,2018PhFl...30d6602K}. This is indicative of
a too weak rotational influence on the flow leading to a too high
Rossby number. Simulations also suggest that convective velocities do
not need to be off by more than 20-30 per cent for the rotation
profile to flip \citep{KKB14}. The discrepancy between simulations and
observations has been dubbed the convective conundrum
\citep{2016AdSpR..58.1475O}.

There are several possibilities why current simulations may
overestimate convective velocity amplitudes. For example, the Rayleigh
numbers in the simulations can be too low \citep[e.g.][]{FH16} or that
convection in the Sun is driven only in thin near-surface layer,
whilst the rest of the CZ is mixed by cool entropy rain emanating from
the near-surface regions \citep{Sp97,Br16}. In such scenarios the bulk
of the CZ would be weakly subadiabatic while the convective flux would
still be directed outward due to a non-local non-gradient contribution
to the convective energy flux \citep{1961JAtS...18..540D,De66}. Such
layers have been named Deardorff zones \citep{Br16}, and have been
detected in many simulations
\citep[e.g.][]{CG92,1993A&A...277...93R,2015ApJ...799..142T,2017ApJ...843...52H,2017ApJ...845L..23K}.
However, the effect of Deardorff layers on velocity amplitudes in
rotating convection in spherical coordinates appears to be weak
\citep{2019GApFD.113..149K}. Furthermore, while there is evidence of
the importance of surface driving \citep[e.g.][]{CR16}, it appears
that this effect is also rather weak \citep{2019SciA....5.2307H}.

Another possible cause of the discrepancy are the unrealistic Prandtl
numbers typically used in simulations. It is numerically convenient to
use $\Pra = 1$, although several recent studies have explored the
possibility that stellar convection might operate in a high effective
(turbulent) Prandtl number regime with $\Pra\gtrsim1$
\citep[e.g.][]{2016AdSpR..58.1475O,2017ApJ...851...74B,2018PhFl...30d6602K}. A
conclusion from these studies is that while the average velocity
amplitude decreases as the Prandtl number increases, turbulent angular
momentum transport is predominantly downward and leads to exacerbation
of the anti-solar differential rotation issue at solar rotation rate
and luminosity \citep{2018PhFl...30d6602K}. Apart from the theoretical
problem of explaining how the Sun switches from $\Pra\ll1$ suggested
by microphysics to $\Pra\gtrsim1$, the numerical studies appear to
disfavour the possibility of an effective Prandtl number greater than
unity in the Sun.

The small Prandtl number limit, which is indicated by estimates of
molecular diffusivities in stellar CZs
\citep[e.g.][]{2019ApJ...876...83A,2020RvMP...92d1001S}, has also been
considered in various studies. An early work by
\cite{1962JGR....67.3063S} explored the limit of zero Prandtl number
in Boussinesq convection. He showed that the temperature fluctuations
are enslaved to vertical motions which leads to highly nonlinear
driving of convection. Subsequent numerical studies in the low-$\Pra$
regime have shown that convection becomes highly inertial with a
tendency for coherent large-scale flow structures to intensify
\citep[e.g.][]{2004PhRvE..69b6302B} along with vigorous
turbulence. Studies of compressible convection have also explored the
Prandtl number dependence, although its signifigance has gone largely
unrecognised. For example, the results of \cite{CBTMH91} show that the
net energy flux due to downflows is reduced as a function of $\Pra$,
whereas the downward kinetic energy flux and upward enthalpy flux both
increase as $\Pra$ decreases. These authors, however, did not connect
this to the changing Prandtl number but rather as a consequence of an
increase in the Reynolds number or the degree of turbulence. Results
similar to those of \cite{CBTMH91} were reported also in
\cite{1993A&A...279..107S}. On the other hand, \cite{BCNS05} found
that the correlation coefficients of velocity and temperature
fluctuations with the enthalpy flux remained $\Pra$-dependent at least
to Reynolds numbers of the order of $10^3$. Finally,
\cite{2018ApJ...856...13O} studied non-rotating anelastic convection
in spherical coordinates and found that the overall velocity amplitude
increases as the Prandtl number decreases while the specral
distribution of velocity is insensitive to $\Pra$.

Here the Prandtl number dependence of convective flow statistics,
overshooting, and energy fluxes are studied systematically with a
hydrodynamic convection setup in Cartesian setup. A motivation of the
current study is a prior work \citep{2019A&A...631A.122K} where it was
found that convective overshooting is sensitive to the Prandtl number
and that earlier numerical results predicting weak overshooting for
solar parameters were obtained from simulations where $\Pra\gtrsim1$
or $\Pra\gg1$. This implies a change in the magnitudes, dominant
scales, or other transport properties, such as correlations with
thermodynamic quantities of convective flows, all of which will be
investigated in the present study. Overshooting is also closely
related to the convective energy transport which is another focus of
the current study. In particular, we will study the overall transport
properties as a function of $\Pra$ and the respective roles of up- and
downflows.

\section{The model} \label{sect:model}

The set-up used in the current study is the same as that in
\cite{2019A&A...631A.122K}. We solve the equations for compressible
hydrodynamics
\begin{eqnarray}
\frac{D \ln \rho}{D t} &=& -\bm\nabla \bm\cdot \uuu, \label{equ:dens}\\
\frac{D\uuu}{D t} &=& {\bm g} -\frac{1}{\rho}(\bm\nabla p - \bm\nabla \bm\cdot 2 \nu \rho \bm{\mathsf{S}}),\label{equ:mom} \\
T \frac{D s}{D t} &=& -\frac{1}{\rho} \left[\bm\nabla \bm\cdot \left(\FFF_{\rm rad} + \FFF_{\rm SGS}\right) + \Gamma_{\rm cool} \right] + 2 \nu \bm{\mathsf{S}}^2,
\label{equ:ent}
\end{eqnarray}
where $D/Dt = \pd/\pd t + \uuu\cdot\bm\nabla$ is the advective
derivative, $\rho$ is the density, $\uuu$ is the velocity,
$\bm{g}=-g\hat{\bm{e}}_z$ with $g>1$ is the acceleration due to
gravity, $p$ is the pressure, $T$ is the temperature, $s$ is the
specific entropy, and $\nu$ is the kinematic viscosity. Furthermore,
$\FFF_{\rm rad}$ and $\FFF_{\rm SGS}$ are the radiative and turbulent
SGS fluxes, respectively, and $\Gamma_{\rm cool}$ describes cooling
near the surface. $\SSt$ is the traceless rate-of-strain tensor with
\begin{eqnarray}
\SStij = \onehalf (u_{i,j} + u_{j,i}) - \onethird \delta_{ij} \bm\nabla\cdot\uuu.
\end{eqnarray}
Radiation is modeled via the diffusion approximation, corresponding to
an optically thick, fully ionized gas. The ideal gas equation of state
$p= (\cP - \cV) \rho T =\calR \rho T$ is assumed, where $\calR$ is the
gas constant, and $c_{\rm P}$ and $c_{\rm V}$ are the specific heats
at constant pressure and volume, respectively. The radiative flux is
given by
\begin{eqnarray}
\FFF_{\rm rad} = -K\bm\nabla T,
\label{equ:Frad}
\end{eqnarray}
where $K(\rho,T)$ is the radiative heat conductivity,
\begin{eqnarray}
K = \frac{16 \sigma_{\rm SB} T^3}{3 \kappa \rho},
\label{equ:Krad1}
\end{eqnarray}
with $\sigma_{\rm SB}$ being the Stefan-Boltzmann constant where
$\kappa$ is the opacity. The latter is assumed to obey a power law
\begin{eqnarray}
\kappa = \kappa_0 (\rho/\rho_0)^a (T/T_0)^b,
\label{equ:kappa}
\end{eqnarray}
where $\rho_0$ and $T_0$ are reference values of density and
temperature. In combination, \Eqsa{equ:Krad1}{equ:kappa} give
\begin{eqnarray}
K(\rho,T) = K_0 (\rho/\rho_0)^{-(a+1)} (T/T_0)^{3-b}.
\label{equ:Krad2}
\end{eqnarray}
With the choices $a=1$ and $b=-7/2$ this corresponds to the Kramers
opacity law \citep{WHTR04}, which was first used in convection
simulations by \cite{2000gac..conf...85B}.

The fixed flux at the bottom of the domain ($\Fbot$) fixes the initial
profile of radiative diffusivity, $\chi=K/(\cP\rho)$, which varies
strongly as a function of height. Thus additional SGS diffusivity is
added in the entropy equation to keep the simulations numerically
feasible. Here the SGS flux is formulated as
\begin{eqnarray}
\FFF_{\rm SGS} = -\rho T \chiSGS \bm\nabla s',
\label{equ:FSGS}
\end{eqnarray}
where $s'=s-\mean{s}$ is the fluctuation of the specific entropy from
the horizontally averaged mean which is denoted by an overbar. The
mean $\mean{s}=\mean{s}(z,t)$ is computed at each timestep. The
coefficient $\chiSGS$ is constant throughout the simulation
domain\footnote{$\chiSGS$ corresponds to $\chiSGSo$ in
  \cite{2019A&A...631A.122K}.}. The net horizontally averaged SGS flux
is negligibly small, such that $\mean{\FFF}_{\rm SGS} \approx 0$.

The cooling near the surface is given by
\begin{eqnarray}
\Gamma_{\rm cool} = - \Gamma_0 h(z) [T_{\rm cool} - T(\xxx,t)],
\label{equ:cool}
\end{eqnarray}
where $\Gamma_0$ is a cooling luminosity, $T=e/c_{\rm V}$ is the
temperature, $e$ is the internal energy, and $T_{\rm cool}=T_{\rm
  top}$ is the fixed reference temperature at the top boundary.

\subsection{Geometry, initial and boundary conditions}

The computational domain is a rectangular box where $z_{\rm bot} \leq
z \leq z_{\rm top}$ is the vertical coordinate. With $z_{\rm
  bot}/d=-0.45$ and $z_{\rm top}/d=1.05$, where $d$ is the depth of
the initially isentropic layer (see below), the vertical extent is
$L_z = z_{\rm top} - z_{\rm bot} = 1.5d$. The horizontal size of the
domain is $L_{\rm H}/d=4$ and the horizontal coordinates $x$ and $y$
run from $-2d$ to $2d$.

The initial stratification consists of three layers such that the two
lower layers are polytropic with polytropic indices $n_1=3.25$
($z_{\rm bot}/d \leq z/d \leq 0$) and $n_2=1.5$ ($0 \leq z/d \leq
1$). The uppermost layer above $z/d=1$ is initially isothermal with
$T=T_{\rm top}$. The latter mimics a photosphere where radiative
cooling is efficient. The initial stratification is set by the
normalized pressure scale height at the top boundary
\begin{eqnarray}
\xi_0 = \frac{{\cal R}T_{\rm top}}{gd}.
\end{eqnarray}
All of the current runs have $\xi_0=0.054$. The choices of $n_1$ and
$n_2$ reflect the expected thermal stratifications in the radiative
(see \cite{BB14} and Appendix~A of \cite{Br16}) and convective layers,
respectively. Prior experience confirms these choices to be valid
\citep[see, e.g.][]{2019GApFD.113..149K}, although the extent of the
CZ is an outcome of the simulation rather than fixed by the input
parameters
\citep[][]{2017ApJ...845L..23K,2019A&A...631A.122K}. Convection ensues
because the system in initially in thermal inequilibrium.

The horizontal boundaries are periodic and on the vertical boundaries
impenetrable and stress-free boundary conditions according to
\begin{eqnarray}
\frac{\pd u_x}{\pd z} = \frac{\pd u_y}{\pd z} = u_z = 0,
\end{eqnarray}
are imposed. The temperature gradient at the bottom boundary is set
according to
\begin{eqnarray}
\frac{\pd T}{\pd z} = -\frac{F_\tbot}{K_\tbot},
\end{eqnarray}
where $F_{\rm bot}$ is the fixed input flux and $K_\tbot(x,y,\zbot,t)$
is the value of the heat conductivity at the bottom of the domain. On
the upper boundary, constant temperature $T=T_{\rm top}$ coinciding
with the initial value is assumed.

\subsection{Units, control parameters, and simulation strategy}

The units of length, time, density, and entropy are given by
\begin{eqnarray}
[x] = d,\ \ \ [t] = \sqrt{d/g},\ \ \ [\rho] = \rho_0,\ \ \ [s] = \cP,
\end{eqnarray}
where $\rho_0$ is the initial value of density at $z=z_{\rm top}$. The
models are fully defined by choosing the values of the kinematic
viscosity $\nu$, gravitational acceleration $g$, the values of $a$,
$b$, $K_0$, $\rho_0$, $T_0$, $\Gamma_0$ and the SGS and effective
Prandtl numbers
\begin{eqnarray}
\PraSGS = \frac{\nu}{\chiSGS},\ \Praeff(z) =
\frac{\nu}{\chiSGS+\chi(z)},
\end{eqnarray}
along with the cooling profile $h(z)$. The values of $K_0$, $\rho_0$,
$T_0$ are subsumed into a new variable $\widetilde{K}_0=K_0
\rho_0^{a+1} T_0^{b-3}$ which is fixed by assuming $\Frad(z_{\rm
  bot})=\Fbot$ in the initial state. The profile $h(z)=1$ for $z/d\geq
1$ and $h(z)=0$ for $z/d<1$, connecting smoothly over a layer of width
$0.025d$. The normalized flux is given by
\begin{eqnarray}
  \mathscr{F}_{\rm n} = \Fbot/\rho_\tbot c_{\rm s,bot}^3,
\end{eqnarray}
where $\rho_\tbot$ and $c_{\rm s,bot}$ are the density and the sound
speed, respectively, at $z_{\rm bot}$ in the initial non-convecting
state. The current runs have $\Fn\approx 4.6\cdot 10^{-6}$
corresponding to runs K3 and K3h in \cite{2019A&A...631A.122K}.

The advective terms in \Equsa{equ:dens}{equ:ent} are formulated in
terms of a fifth-order upwinding derivative with a hyperdiffusive
sixth-order correction with a flow-dependent diffusion coefficient,
see Appendix~B of \cite{DSB06}.

\subsection{Diagnostics quantities}

The following quantities are outcomes of the simulations that can only
be determined a posteriori. These include the global Reynolds number
and the SGS and effective P\'eclet numbers
\begin{eqnarray}
\Rey\!=\!\frac{\urms}{\nu k_1},\ \Pe_{\rm SGS}\!=\!\frac{\urms}{\chi_{\rm SGS} k_1},\ \Pe(z)\!=\!\frac{\urms}{[\chi_{\rm SGS}+\chi(z)] k_1},
\end{eqnarray}
where $\urms$ is the volume averaged rms-velocity and $k_1=2\pi/d$ is
an estimate of the largest eddies in the system.

To assess the level of supercriticality of convection the Rayleigh
number is defined as:
\begin{eqnarray}
\Ra(z) &=& \frac{gd^4}{\nu [\chiSGS+\chi(z)]}\left( - \frac{1}{\cP}\frac{{\rm d}s}{{\rm d}z} \right).
\end{eqnarray}
The Rayleigh number varies as a function of height and is quoted near
the surface at $z/d=0.85$ for all models. Conventionally the Rayleigh
number in the hydrostatic, non-convecting, state is one of the control
parameters. In the current models with Kramers conductivity the
convectively unstable layer in the hydrostatic case is very thin and
confined to the near-surface layers \citep{Br16}. Thus the Rayleigh
numbers are quoted from the thermally saturated statistically
stationary states.

\begin{table}[t!]
\centering
\caption[]{Summary of the runs.}
  \label{tab:runs1}
      $$
          \begin{array}{p{0.06\linewidth}cccrrrrcr}
          \hline
          \hline
          \noalign{\smallskip}
          Run  & \PraSGS  & \Pra_{\rm eff}^{\rm top}  & \Pra_{\rm eff}^{\rm bot}  & \Rey  & \Pe_{\rm SGS}  & \Pe^{\rm top}  & \Pe^{\rm bot}  & \Ra \\
          \hline
          \hline
          A01     &    0.1  &    0.1  &    0.1  &   48  &    4.8  &   4.8  &    4.7  &  5.9 \cdot 10^{5} \\
          B01     &    0.1  &    0.1  &    0.1  &   93  &    9.3  &   9.3  &    9.2  &  2.3 \cdot 10^{6} \\
          C01     &    0.1  &    0.1  &    0.1  &  190  &     19  &    19  &     18  &  8.4 \cdot 10^{6} \\
          C01b    &    0.1  &    0.1  &    0.1  &  318  &     32  &    32  &     30  &  2.3 \cdot 10^{7} \\
          C01c    &    0.1  &    0.1  &    0.1  &  473  &     47  &    47  &     43  &  5.2 \cdot 10^{7} \\
          \hline
          A02     &    0.2  &    0.2  &    0.2  &   44  &    8.8  &   8.8  &    8.6  &  9.7 \cdot 10^{5} \\
          B02     &    0.2  &    0.2  &    0.2  &   89  &     18  &    18  &     17  &  4.0 \cdot 10^{6} \\
          C02     &    0.2  &    0.2  &    0.2  &  178  &     36  &    35  &     33  &  1.6 \cdot 10^{7} \\
          \hline
          A05     &    0.5  &    0.5  &    0.5  &   41  &     20  &    20  &     20  &  2.1 \cdot 10^{6} \\
          B05     &    0.5  &    0.5  &    0.5  &   85  &     42  &    42  &     39  &  9.9 \cdot 10^{6} \\
          \hline
          A1      &    1.0  &    1.0  &    0.9  &   39  &     39  &    39  &     35  &  4.2 \cdot 10^{6} \\
          B1      &    1.0  &    1.0  &    0.8  &   84  &     84  &    83  &     70  &  2.1 \cdot 10^{7} \\
          C1      &    1.0  &    1.0  &    0.7  &  175  &    175  &   170  &    127  &  8.1 \cdot 10^{7} \\
          \hline
          A2      &    2.0  &    2.0  &    1.7  &   38  &     75  &    74  &     63  &  8.4 \cdot 10^{6} \\
          B2      &    2.0  &    1.9  &    1.4  &   82  &    164  &   159  &    118  &  4.1 \cdot 10^{7} \\
          \hline
          A5      &    5.0  &    4.8  &    3.3  &   36  &    181  &   175  &    121  &  2.1 \cdot 10^{7} \\
          B5      &    5.0  &    4.7  &    2.5  &   80  &    399  &   374  &    203  &  9.9 \cdot 10^{7} \\
          \hline
          A10d    &     10  &    9.9  &    9.1  &  2.7  &     27  &    27  &     24  &  3.5 \cdot 10^{5} \\
          A10c    &     10  &    9.9  &    8.3  &  5.9  &     59  &    59  &     49  &  1.0 \cdot 10^{6} \\
          A10b    &     10  &    9.6  &    6.7  &   16  &    163  &   157  &    109  &  8.5 \cdot 10^{6} \\
          A10     &     10  &    9.3  &    5.0  &   35  &    354  &   331  &    178  &  4.1 \cdot 10^{7} \\
          B10     &     10  &    8.8  &    3.4  &   79  &    791  &   697  &    269  &  1.9 \cdot 10^{8} \\
          C10     &     10  &    7.8  &    2.1  &  167  &   1667  &  1308  &    347  &  6.4 \cdot 10^{8} \\
          \hline
          \end{array}
          $$ \tablefoot{The superscripts top and bot refer to
            $z/d=0.85$ and to the bottom of the CZ, $\zcz$,
            respectively. $\Gamma_0 = 2.5 \cP (g\rho_0)^{1/2}$ in all
            of the runs. The grid resolutions are $288^3$ (set A),
            $576^3$ (B), and $1152^3$ (C), respectively.}
\end{table}

Contributions to the horizontally averaged vertical energy flux are:
\begin{eqnarray}
\mFrad  &=& - \mean{K} \frac{\pd \mean{T}}{\pd z},\\
\mFenth &=& \cP \mean{(\rho u_z)' T'}, \label{equ:Fenth}\\
\mFkin  &=& \onehalf \mean{\rho \uuu^2 u_z'}, \label{equ:Fkin} \\
\mFvisc &=& -2 \nu \mean{\rho u_i \mathsf{S}_{iz}}\\
\mFcool &=& \int_{z_{\rm bot}}^{z_{\rm top}} \Gamma_{\rm cool} {\rm d}z.
\end{eqnarray}
Here the primes denote fluctuations and overbars horizontal
averages. The
total convected flux \citep{CBTMH91} is the sum of the enthalpy and
kinetic energy fluxes:
\begin{eqnarray}
\mFconv = \mFenth + \mFkin. \label{equ:convflux}
\end{eqnarray}
The CZ is defined as the region where $\mFconv > 0$. The vertical
position of the bottom of the CZ is given by $\zcz$. Error estimates
for diagnostics are obtained by dividing the time series in three
equally long parts and computing averages over each of them. The
largest deviation of these averages from the time average over the
whole time series is taken to represent the error.

The {\sc Pencil Code}
\citep{2021JOSS....6.2807P}\footnote{\href{https://github.com/pencil-code/}{https://github.com/pencil-code/}}
was used to produce the simulations. At the core of the code is a
switchable finite difference solver for partial differential equations
that can be used to study a wide selection of physical problems. In
the current study a third-order Runge-Kutta time stepping method and
centered sixth-order finite differences for spatial derivatives were
used \citep[cf.][]{B03}.

\begin{figure*}
  \begin{center}
  \includegraphics[width=0.8\textwidth]{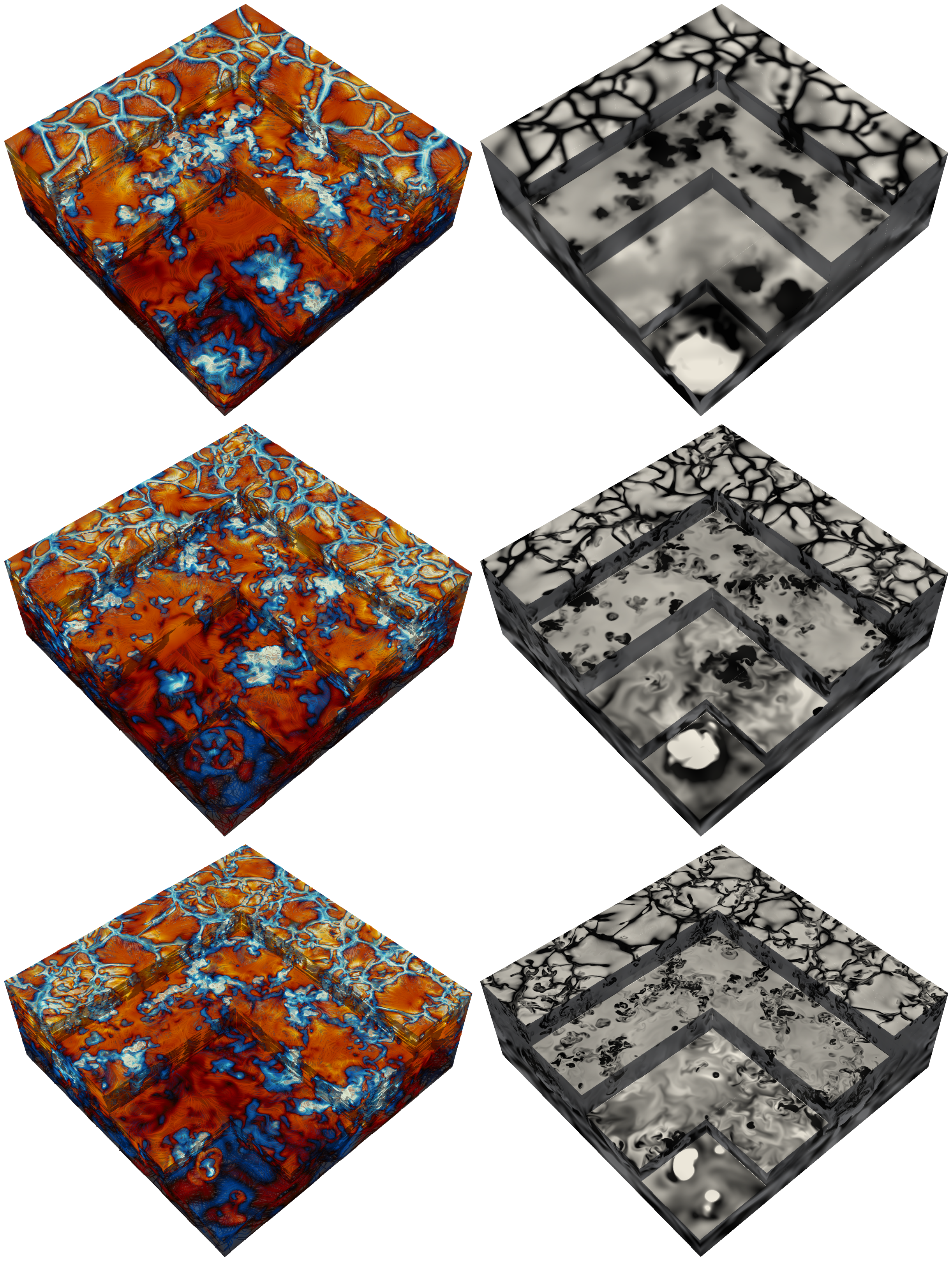}
  \end{center}
  \vspace{-.5cm}
  \caption{Left panels: Normalized vertical velocity $\tilde{u}_z=u_z
    (dg)^{-1/2}$ (colour contours) and streamlines of the flow from
    runs C01 ($\PraSGS=0.1$, top panel), C1 ($\PraSGS=1$, middle), and
    C10 ($\PraSGS=10$, bottom). The colours of the streamlines
    indicate the local vertical velocity.  Right panels: normalized
    entropy fluctuations $\tilde{s}'(\xxx) = [s(\xxx) -
      \mean{s}(z)]/s'_{\rm rms}(z)$ from the same runs.}
\label{fig:pboxes_ad1152a3}
\end{figure*}

\section{Results} \label{sect:results}

All of the runs discussed in the present study were branched off from
run K3h of \cite{2019A&A...631A.122K}, see \Table{tab:runs1} for a
summary of the simulations. A thermally saturated snapshot of this run
was used to produce new low resolution models, labeled the A set, at
SGS Prandtl numbers ranging between 0.1 and 10. The runs in the
intermediate (high) resolution set B (C) were remeshed from saturated
snapshots of the corresponding runs in A (B) set. Only a subset of
$\PraSGS$ values were done at the highest grid resolution in set C.

\subsection{Flow characteristics}

\Figu{fig:pboxes_ad1152a3} shows visualizations of the flows and
entropy fluctuations in cases with $\PraSGS=0.1$, $1$, and $10$
($\Rey=190$, $175$, and $167$) corresponding to runs C01, C1, and
C10. Visual inspection of the velocity patterns does not reveal
notable differences at large scales such that the dominant granule
sizes in the three cases are similar. Smaller scale structures in the
velocity appear especially near the surface as the SGS Prandtl number
increases; see the middle and bottom panels of
\Figa{fig:pboxes_ad1152a3}. Nevertheless, the velocity patterns are
remarkably similar in comparison to the entropy fluctuations which
change dramatically as $\PraSGS$ increases from 0.1 to 10. In the
$\PraSGS=0.1$ case the strong downflows coincide with smooth regions
of negative (cool) entropy fluctuations, whereas the surrounding areas
are almost featureless. In contrast, in the $\PraSGS=10$ case the
smooth negative entropy regions have disintegrated into numerous
smaller structures that are often detached from each other. The
entropy fluctuations in the $\PraSGS=1$ run shows an intermediate
behavior with traces of both large- and small-scale structures.

\begin{figure}
  \includegraphics[width=\columnwidth]{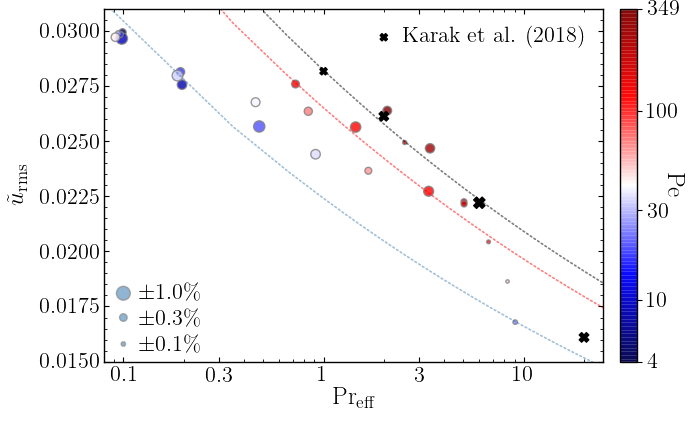}
  \caption{Volume and time averaged rms velocity as a function
    $\Praeff(\zcz)$ for all of the current runs (circles). The colours
    (sizes) of the symbols indicate the P\'eclet number (relative
    error) as indicated in the colour bar (legend). The dotted lines
    show power laws proportional to $\Praeff^{-0.13}$ for
    reference. The crosses show the scaled results from the
    non-rotating runs R0P[1,2,6,20] of \cite{2018PhFl...30d6602K}; see
    their Table~1.}
\label{fig:purms}
\end{figure}

The averaged rms velocity as a function $\Praeff$ from all runs is
shown in \Figa{fig:purms}. There is a tendency for $\urms$ to increase
with decreasing $\PraSGS$ which was first reported by
\cite{1993A&A...279..107S} in compressible convection. Data for
approximately constant $\Pe$ is suggestive of a power law with
exponent around $-0.13$, see the dotted lines in \Figa{fig:purms}. A
similar dependence can be seen in the suitably scaled data of
non-rotating spherical shell simulations R0P[1,2,6] of
\cite{2018PhFl...30d6602K}, see the crosses in
\Figa{fig:purms}. However, their run with the highest $\Pra$ (R0P20)
appears to be an outlier. The results of \cite{2018ApJ...856...13O}
also indicate an increase of kinetic energy as the Prandtl number
decreases. Furthermore, qualitatively similar results have been
reported from Boussinesq convection
\citep[e.g.][]{2004PhRvE..69b6302B,2016JFM...802..147S}, although
there the dependence of $\urms$ on $\Pra$ is much steeper. This is
because in Rayleigh-B\'enard convection the total flux transported by
convection depends strongly on the Prandtl number.

\begin{figure}
  \includegraphics[width=\columnwidth]{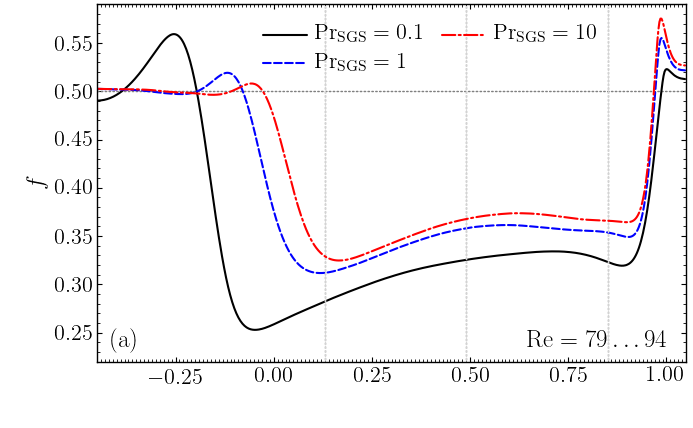}
  \includegraphics[width=\columnwidth]{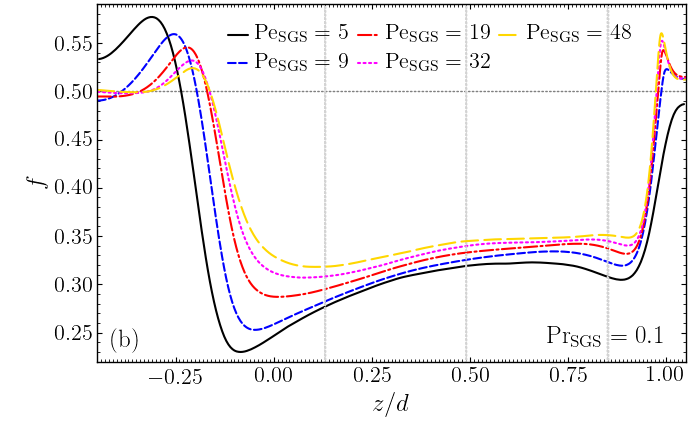}
\caption{Filling factor of downflows for SGS Prandtl numbers 0.1
  (solid line), 1 (dashed), and 10 (dash-dotted) as indicated by the
  legend from runs B01, B1, and B10 \emph{(a)}. Filling factor in runs
  with $\PraSGS=0.1$ with different $\Pe_{\rm SGS}$ \emph{(b)}. The
  vertical dotted lines indicate the depths from which representative
  data for the top, middle, and bottom of the CZ are considered.}
\label{fig:fillingfactor}
\end{figure}

A distinct characteristic of convection is that the vertical flows are
the main transporter of the energy flux. A diagnostic of the structure
of vertical flows is the filling factor $f$ of up- or downflows. Here
the filling factor is defined as the area the downflows occupy at each
depth such that the horizontally averaged vertical velocity us given
by:
\begin{eqnarray}
\mean{u}_z(z) = f(z) \mean{u}_z^\downarrow(z) + [1 - f(z)] \mean{u}_z^\uparrow(z).
\end{eqnarray}
where $\mean{u}_z$ is the total vertical velocity whereas
$\mean{u}_z^\downarrow$ and $\mean{u}_z^\uparrow$, respectively, are
the mean velocities in the down- and
upflows. \Figu{fig:fillingfactor}(a) shows $f$ for runs B01, B1, and
B10 with $\PraSGS=0.1$, $1$, and $10$. These results indicate that the
filling factor decreases with decreasing SGS Prandtl number. However,
the change is relatively minor such that $f$ differs by roughly 20 per
cent in the bulk of the CZ between the extreme cases with
$\PraSGS=0.1$ and $10$. The filling factor plays an important role in
analytic and semi-analytic two-stream models of convection
\citep[e.g.][]{2004ApJ...607.1046R,Br16}. For example, in the updated
mixing length model of \cite{Br16}, a very small filling factor is
needed in cases where the Schwarzschild unstable part of the CZ is
particularly shallow. The current simulations suggest that the filling
factor goes to that direction when $\PraSGS$ decreases, but it appears
that the smallest values of the order of $10^{-4}$ in some of the
models of \cite{2004ApJ...607.1046R} and \cite{Br16} are ruled out.

The filling factor increases as a function of $\Rey$ and $\Pe_{\rm
  SGS}$ for a given $\PraSGS$, see \Figa{fig:fillingfactor}(b) for
representative results for $\PraSGS=0.1$ where $\chiSGS \gg \chi$. It
is plausible that a growing contribution from nearly isotropic
small-scale eddies that become more prominent with increasing $\Rey$
is partly responsible for the increasing trend of $f$ as a function of
the Reynolds number. However, preliminary studies using low-pass
filtered data suggest that this effect is very subtle. Irrespective of
the $\Rey$ dependence of $f$, the energetics of the underlying larger
scale granulation appear to be almost unaffected by $\Rey$; see
\Sec{sec:flowstat}.

\subsection{Overshooting below the convection zone}

It is of interest to study the extent of convective overshooting below
the CZ given the systematic dependence of the overall convective
velocities on $\PraSGS$. The same definition of overshooting as in
\cite{2019A&A...631A.122K} is used. This is outlined by defining the
bottom of the CZ ($\zcz$) to be the depth where $\mFconv$ changes from
positive to negative. This depth is used to obtain a reference value
of the horizontally averaged kinetic energy flux $\mFkin^{\rm
  ref}=\mFkin(\zcz,t)$. The instantaneous overshooting depth $\zos$ is
then taken to be the depth where $\mFkin(z,t)$ drops below
$10^{-2}\mFkin^{\rm ref}$. The mean thickness of the overshoot layer
is defined as
\begin{eqnarray}
d_{\rm os} = \frac{1}{\Delta t}\int_{t_0}^{t_1} [\zcz(t) - \zos(t)] dt,
\end{eqnarray}
where $\Delta t = t_1 - t_0$, where $t_0$ and $t_1$ denote the
beginning and end of the time averaging period.

\begin{figure}
  \includegraphics[width=\columnwidth]{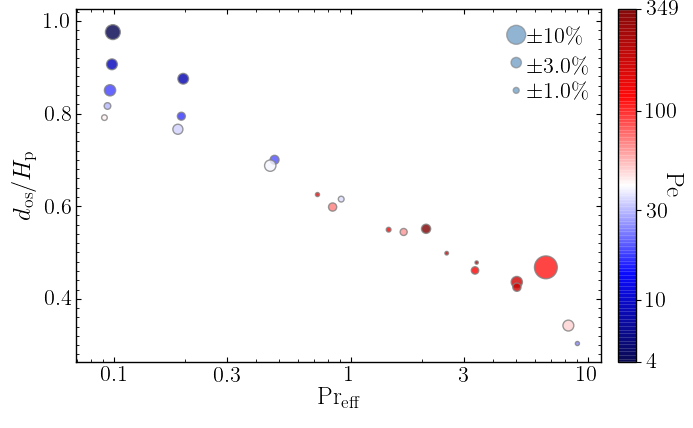}
  \caption{Thickness of the overshoot layer $d_{\rm os}$ normalized by
    the pressure scale height $H_p(\zcz)$ for all runs as a function
    of $\Praeff(\zcz)$. The colours (sizes) of the symbols indicate
    the P\'eclet number (relative error) as indicated in the colour
    bar (legend).}
\label{fig:poshoot}
\end{figure}

\Figu{fig:poshoot} shows $d_{\rm os}$ from all runs as a function of
$\PraSGS$. The current results confirm the conjecture of
\cite{2019A&A...631A.122K} that $d_{\rm os}$ is sensitive to the
Prandtl number. However, the results still depend of the P\'eclet
number especially for low $\PraSGS$ where it is challenging to reach
high values of $\Pe$. Nevertheless, comparing $d_{\rm os}$ for
approximately the same $\Pe$, for example $\Pe\approx 40$ (light blue
symbols in \Figa{fig:poshoot}), shows that the overshooting is
increasing monotonically as $\PraSGS$ decreases. The difference of
$d_{\rm os}$ between the cases $\PraSGS=1$ and $\PraSGS=0.1$ is about
30 per cent. However, the difference is decreasing as the P\'eclet
number increases but it does not appear likely that the $\Pra$
dependence would disappear at even higher P\'eclet numbers.

\begin{figure}
  \includegraphics[width=\columnwidth]{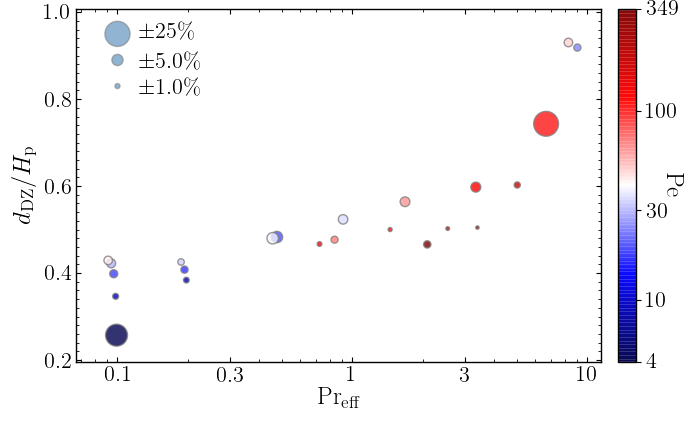}
  \caption{Thickness of the Deardorff layer $d_{\rm DZ}$ normalized by
    the pressure scale height $H_p(\zcz)$ for all runs as a function
    of $\Praeff(\zcz)$. The colours (sizes) of the symbols indicate
    the P\'eclet number (relative error) as indicated in the colour
    bar (legend).}
\label{fig:pDeardorff}
\end{figure}

Several numerical studies have shown that the deep parts of density
stratified CZs are often weakly stably stratified
\citep[e.g.][]{CG92,2015ApJ...799..142T,2017ApJ...851...74B,2017ApJ...843...52H,2017ApJ...845L..23K}. Such
layers have also been found from semi-global and global simulations of
rotating solar-like CZs
\citep[e.g.][]{2018PhFl...30d6602K,2019GApFD.113..149K,2021A&A...645A.141V}
as well as from fully convective spheres \citep{2020arXiv201201259K}.
We call this layer the Deardorff zone after \cite{Br16}. This layer is
characterised by a positive vertical gradient of entropy, $ds/dz>0$,
along with a positive convective energy flux, $\mFconv > 0$. The mean
thickness of the Deardorff zone is defined similarly as $d_{\rm os}$
via
\begin{eqnarray}
d_{\rm DZ} = \frac{1}{\Delta t}\int_{t_0}^{t_1} [\zbz(t) - \zcz(t)] dt,
\end{eqnarray}
where $\zbz$ is the depth where the entropy gradient changes sign.
The results for $d_{\rm DZ}$ are shown in \Figa{fig:pDeardorff}. At
first glance, $d_{\rm DZ}$ shows an opposite trend in comparison to
$d_{\rm os}$ such that it decreases with the Prandtl number but the
dependence on P\'eclet and Reynolds numbers is again strong,
particularly for low $\PraSGS$. However, considering only the largest
$\Pe$ for each $\Praeff$, $d_{\rm DZ}$ is roughly constant around
$0.45H_p$ for $\Praeff \lesssim 3$.

\subsection{Flow statistics and spectral distribution}
\label{sec:flowstat}

\begin{figure*}
  \includegraphics[width=\textwidth]{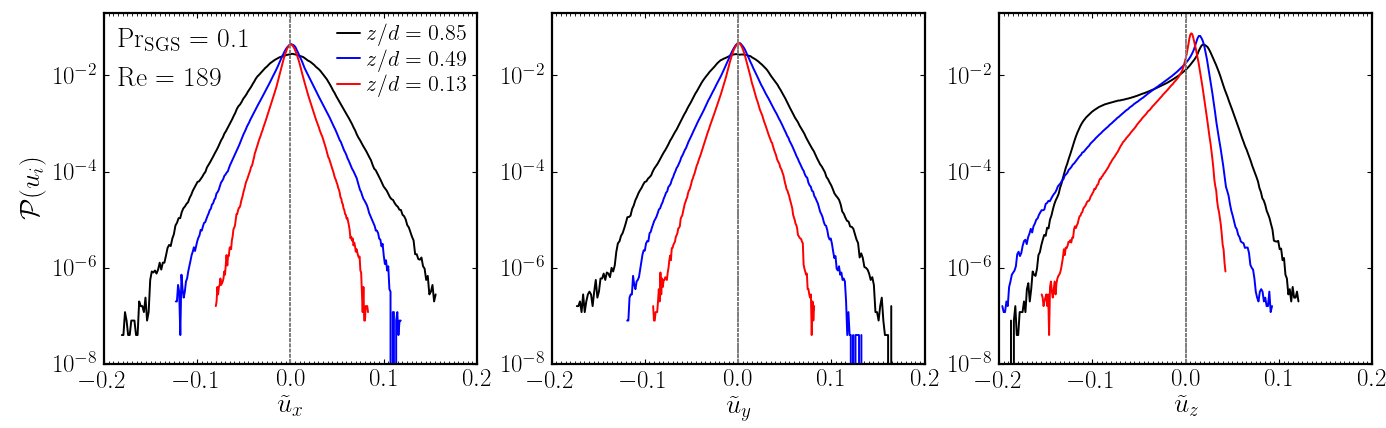}
\caption{Normalized PDFs of $u_x$ (left panel), $u_y$ (middle), and
  $u_z$ (right) from near the surface (black lines), middle (blue) and
  bottom (red) of the CZ as indicated by the legend for run C01 with
  $\Rey=190$ and $\PraSGS=0.1$.}
\label{fig:plot_pdf_ad1152a3}
\end{figure*}

A standard diagnostics in flow statistics is the probability density
function (PDF) which is defined by:
\begin{eqnarray}
\int {\cal P}(u_i) du_i = 1.
\end{eqnarray}
The moments of the PDF carry information about the statistical
properties of the flow. Horizontally averaged instantaneous moments
are given by:
\begin{eqnarray}
{\cal M}^n(u_i,z)\!=\!\frac{1}{L_{\rm H}^2} \int_{-L_{\rm H}/2}^{-L_{\rm H}/2} \int_{-L_{\rm H}/2}^{-L_{\rm H}/2} [u_i(\xxx) - \mean{u}_i(z)]^2 dx dy.
\end{eqnarray}
Here we study the skewness ${\cal S}$ and the kurtosis ${\cal K}$ of
the velocity field:
\begin{eqnarray}
{\cal S} = \frac{{\cal M}^3}{\sigma_u^3},\ \ {\cal K} = \frac{{\cal M}^4}{\sigma_u^4},\ \ \mbox{where}\ \ \sigma_u = ({\cal M}^2)^{1/2}.
\end{eqnarray}
Representative PDFs of the velocity components near the surface, at
the middle and near the base of the CZ from run C01 are shown in
\Figa{fig:plot_pdf_ad1152a3}. The data is obtained by time-averaging
over several snapshots. The horizontal flows have zero means and they
have symmetric distributions around the mean. The vertical flows, on
the other hand, show a bimodal distribution corresponding to the
characteristic up- and downflow structure of convective granulation
which is particularly clear near the surface. Similar results have
been reported from a number of earlier numerical studies in different
contexts and set-ups
\citep[e.g.][]{BJNRST96,2008ApJ...673..557M,HRY15b}.

\begin{figure*}
  \includegraphics[width=\textwidth]{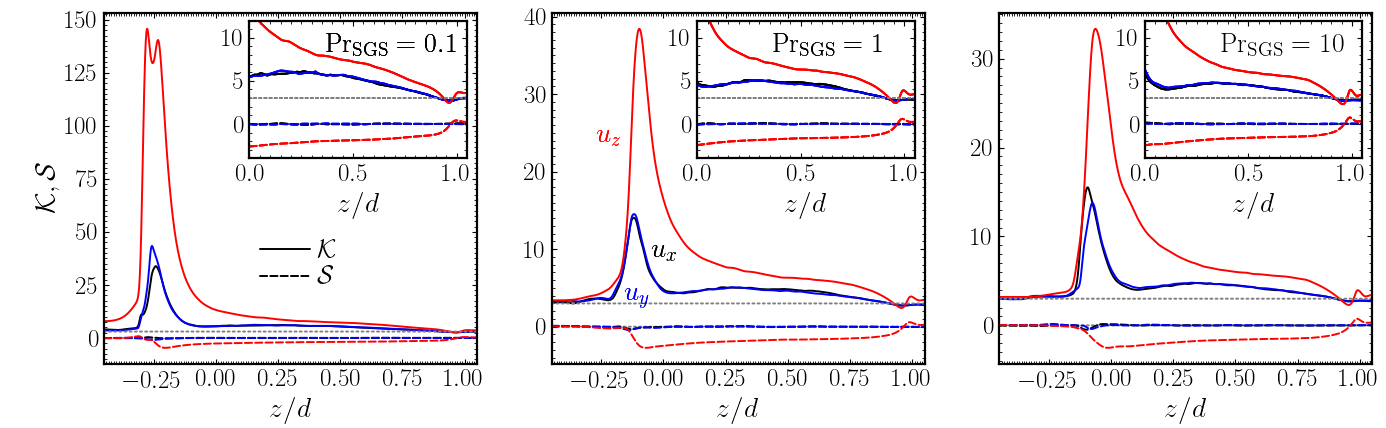}
\caption{Kurtosis ${\cal K}$ (solid lines) and skewness ${\cal S}$
  (dashed) of the velocity components from runs C01 ($\PraSGS=0.1$,
  left panel), C1 ($\PraSGS=1$, middle), and C10 ($\PraSGS=10$,
  right).}
\label{fig:plot_kurtskew_ad1152a3}
\end{figure*}

\Figu{fig:plot_kurtskew_ad1152a3} shows ${\cal S}$ and ${\cal K}$ as
functions of depth from runs C01, C1, and C10. The skewness of the
horizontal flows is essentially zero in all of the cases which is
expected because no systematic horizontal anisotropy is
present. ${\cal S}$ is consistently negative and decreasing with depth
for vertical flows within the CZ. This is indicative of a growing
difference between the statistics of up- and downflows in the deeper
parts. The kurtosis indicates a nearly Gaussian distribution with
${\cal K}\approx3$ for both vertical and horizontal flows near the
surface ($z/d\approx 0.9$). However, ${\cal K}$ increases as a
function of depth such that ${\cal K} \approx 5$ for $u_x$ and $u_y$,
and greater than ten for $u_z$ at the base of the CZ indicating strong
non-Gaussianity. The skewness and kurtosis do not change significantly
in the CZ as a function of $\PraSGS$. The results for ${\cal S}$ and
${\cal K}$ are in agreement with those of \cite{HRY15b} and similar to
the rotating spherical shell simulations of \cite{2008ApJ...673..557M}
notwithstanding the horizontal anisotropy in the latter. Finally,
${\cal K}$ obtains very high values in the overshoot regions
especially for low $\PraSGS$, see left panel of
\Figa{fig:plot_kurtskew_ad1152a3}. This is most likely due to the
highly intermittent turbulence due to the limited number of deeply
penetrating plumes in these regions.

Next we study the velocity amplitudes as functions of spatial scale
from power spectra of the velocity field,
\begin{eqnarray}
E_{\rm K}(k,z,t) &=& \frac{1}{2} \sum_k |\hat{\bm u} (k,z,t)|^2,
\end{eqnarray}
where $k=\sqrt{k_x^2+k_y^2}$ is the horizontal wavenumber and the hat
denotes a two-dimensional Fourier transform. The power spectra are
computed from between ten to 40 snapshots depending on the run and
then time-averaged. Furthermore, normalization is applied such that
\begin{eqnarray}
\tilde{E}_{\rm K}(k,z) &=& \frac{1}{\Delta T}\int_{T_0}^{T_1} \frac{E_{\rm K}(k,z,t)}{\sum_0^{k_{\rm grid}} E_{\rm K}(k,z,t)}dt,
\end{eqnarray}
where $k_{\rm grid} = nx_h/2$ is the Nyquist scale of the horizontal
grid with $nx_h$ grid points. With this normalization the differences
in the shape of the spectra are highlighted whereas the differences in
the absolute magnitude are hidden. This is justified here because we
are currently interested in the effects of the Prandtl number on the
distribution of power as a function of spatial scale.

\begin{figure*}
  \includegraphics[width=\textwidth]{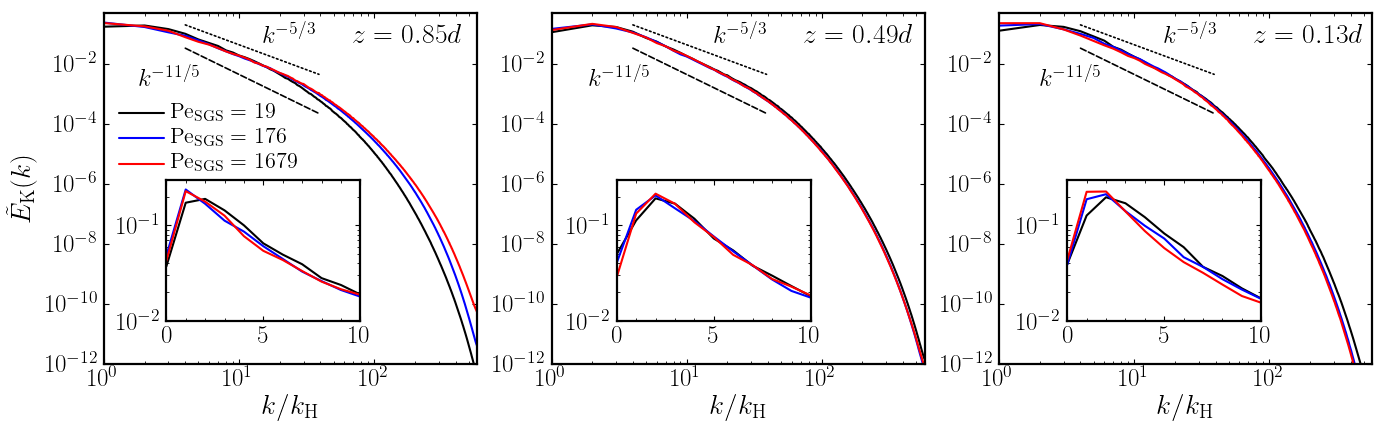}
\caption{Normalized power spectra of velocity from depths $z=0.85d$
  (left panel), $z=0.49d$ (middle), and $z=0.13d$ (right) for
  $\PraSGS=0.1$ (black lines), $1$ (blue), and $10$ (red) and
  $\Rey=167\ldots190$ corresponding to runs C01, C1, and C10,
  respectively. The insets show the low wavenumber part of the
  spectra. The dashed and dotted lines respectively indicate the
  Bolgiano-Obukhov $k^{-11/5}$ and Kolmogorov-Obukhov $k^{-5/3}$
  scalings for reference.}
\label{fig:plot_uu_spectra_ad1152a3}
\end{figure*}

Representative results for $\PraSGS=0.1$, $1$, and $10$ are shown in
\Figa{fig:plot_uu_spectra_ad1152a3} from runs C01, C1, and C10. The
differences in $E_K$ are small and most clearly visible at high
wavenumbers near the surface and near the base of the CZ. The power at
largest scales ($k/k_{\rm H}<3$) is similar in all cases and the
clearest differences are seen at the base of the CZ although even
these features are not particularly pronounced. It is possible that
the horizontal extent of the domain is too small to capture the
largest naturally excited scales because the peak of the power
spectrum is always near the box scale. The spectra show a scaling that
is consistently steeper than the Kolmogorov-Obukhov $k^{-5/3}$
spectrum. The Bolgiano-Obukhov scaling
\citep{1959JGR....64.2226B,1959DoSSR..125..1246O} with $k^{-11/5}$ is
more compatible with the data especially at large $\Rey$; see also
\Figa{plot_uu_spectra_PrSGS01}. However, the inertial range even in
the current highest resolution simulations is very limited and the
conclusions regarding scaling properties are quite
uncertain. Furthermore, there is a puzzling spread of scaling
exponents from convection simulations: early studies with modest
inertial ranges suggested a $k^{-5/3}$ scaling
\citep[e.g.][]{CBTMH91,BJNRST96,2000ApJS..127..159P} whereas more
recent studies \citep[e.g.][]{HRY15b,FH16} suggest clearly shallower
scalings. On the other hand, power spectra of solar surface convection
suggest significantly steeper \citep[][and references
  therein]{2020A&A...644A..44Y} scaling.

\begin{figure*}
  \includegraphics[width=\textwidth]{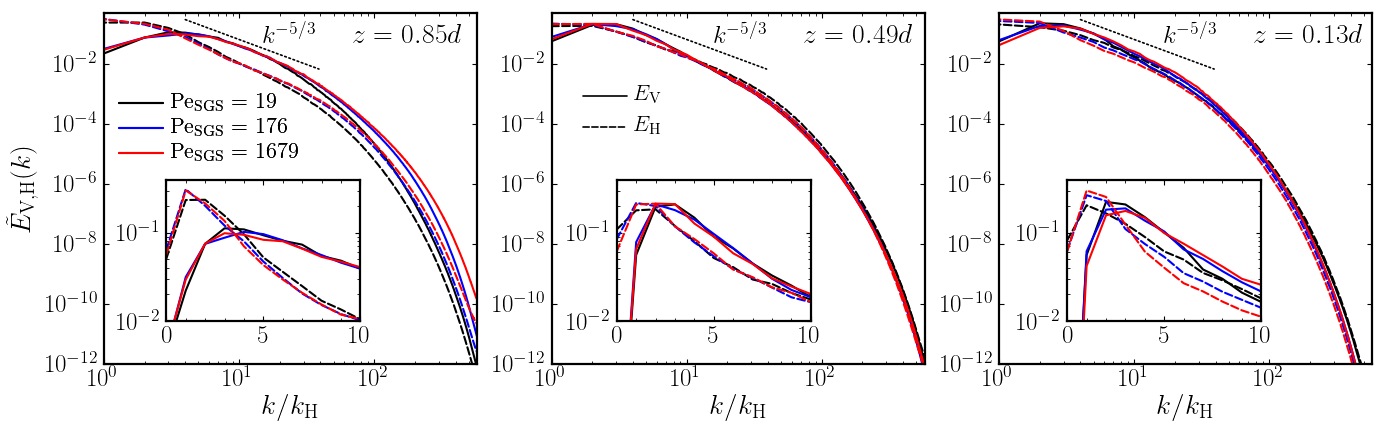}
\caption{Normalized power spectra of vertical ($E_{\rm V}$, solid
  lines) and horizontal velocities $E_{\rm H}$, dashed) from the same
  depths and runs as in \Figa{fig:plot_uu_spectra_ad1152a3}. The inset
  shows low wavenumber contributions.}
\label{fig:plot_uui_spectra_ad1152a3}
\end{figure*}

\begin{figure*}
  \includegraphics[width=\textwidth]{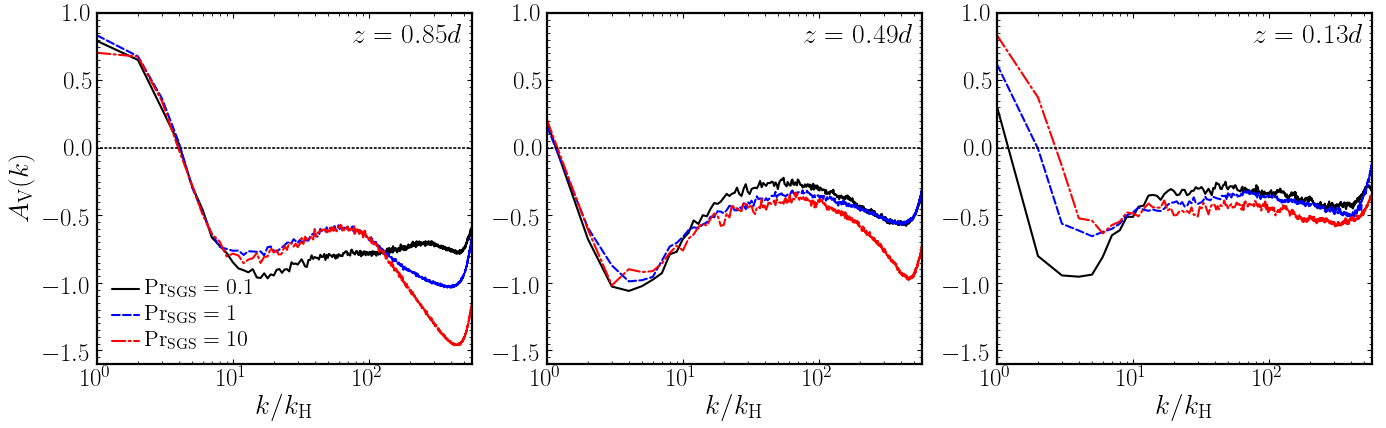}
\caption{Spectral vertical anisotropy parameter $A_{\rm V}(k)$
  according to \Eq{equ:anisok} from near the surface ($z=0.85d$, left
  panel), middle ($z=0.49d$, middle), and near the bottom of the CZ
  ($z=0.13d$, right) for the same runs as in
  \Figa{fig:plot_uu_spectra_ad1152a3}.}
\label{fig:paniso_ad1152a3}
\end{figure*}

To study the anisotropy of the flow, the power spectra of vertical and
horizontal velocities are defined as
\begin{eqnarray}
\int_0^{k_{\rm max}} E_{\rm V}(k,z) dk &=& \onehalf \mean{u_z^2}(z), \\ 
\int_0^{k_{\rm max}} E_{\rm H}(k,z) dk &=& \onehalf \left[\mean{u_x^2}(z) + \mean{u_y^2}(z) \right].
\end{eqnarray}
The same averaging and normalization as above are applied here.
Representative results from the same runs as in
\Figa{fig:plot_uu_spectra_ad1152a3} are shown in
\Figa{fig:plot_uui_spectra_ad1152a3}. The horizontal and vertical
velocity power spectra at all depths indicate a dominance of
horizontal flows at large scales ($k\lesssim 3$) whereas vertical
flows are dominant for larger $k$. The scaling of $E_{\rm H}$ is
consistently steeper than the Kolmogorov-Obukhov $k^{-5/3}$
dependence.

The differences between runs are, however, again small with the
exception of the bottom of the CZ ($z=0.13d$) where a reduction of
$E_{\rm H}$ at large scales $k/k_{\rm H}\leq4$ for $\PraSGS=0.1$ is
seen. The changes in the spectra are rather subtle and an alternative
way to study the spectral distribution of velocity is to consider the
spectral anisotropy parameter $A_{\rm V}$ which is defined as
\citep[][]{2019AN....340..744K}:
\begin{eqnarray}
A_{\rm V}(k,z) \equiv \frac{E_{\rm H}(k,z)-2 E_{\rm V}(k,z)}{E_{\rm K}(k,z)}.\label{equ:anisok}
\end{eqnarray}
Results for $A_{\rm V}$ for the same runs as in
Figs.~\ref{fig:plot_uu_spectra_ad1152a3} and
\ref{fig:plot_uui_spectra_ad1152a3} are shown in
\Figa{fig:paniso_ad1152a3}. Near the surface the large scales are
dominated by horizontal flows such that $A_{\rm V}(k)>0$ for
$k<4$. The large-scale anisotropy for $k/k_{\rm H}<10$ is essentially
identical for the three simulations shown in
\Figa{fig:paniso_ad1152a3}. The run with the highest $\PraSGS$ starts
to deviate from the other two for $k/k_{\rm H}>10$, and the two
remaining runs deviate for $k/k_{\rm H}\gtrsim 150$. Given that the
energy transport is dominated by scales for which $k/k_{\rm H}
\lesssim 30$ (see below), it is likely that the differences of $A_{\rm
  V}$ at large $k/k_{\rm H}$ are not of great importance. The
anisotropy at the middle of the CZ is remarkably similar for all three
cases such that significant deviations occur only for $k/k_{\rm
  H}\gtrsim 100$, see the middle panel of
\Figa{fig:paniso_ad1152a3}. The situation changes dramatically at the
base of the CZ: while $A_{\rm V}$ is essentially the same for all
Prandtl numbers for $k/k_{\rm H}>8$, the results systematically
deviate at larger scales. That is, the flow at largest scales
($k/k_{\rm H}=1$) continues to be horizontally dominated for all
Prandtl numbers but $A_{\rm V}$ is decreasing monotonically with
$\PraSGS$ such the for $\PraSGS=0.1$, $A_{\rm V}$ change of sign from
positive to negative already for $k/k_{\rm H}>1$. This is reflecting
the stronger downflows and deeper overshooting in the low-$\PraSGS$
regime.

\Figu{plot_uu_spectra_PrSGS01} shows normalized velocity power spectra
compensated by $k^{11/5}$ from five runs with $\PraSGS=0.1$ where the
Reynolds number varies between 48 and 473. This figure shows that the
scaling of the velocity spectra for the highest Reynolds numbers at
low $\PraSGS$ is close to or even steeper than the Bolgiano-Obukhov
$k^{-11/5}$ scaling. This is particularly clear at the middle and at
the base of the CZ while no clear scaling can be discerned near the
surface. Evidence for the Bolgiano-Obukhov scaling for the kinetic
energy spectrum has previously been reported from simulations
Rayleigh-B\'enard convection \citep[e.g.][]{2002PhRvE..66a6304C} as
well as from corresponding shell models
\citep{1992PhRvL..69..605B}. However, the generality of the results
for the spectra remain in question as stated earlier.

\begin{figure*}
  \includegraphics[width=\textwidth]{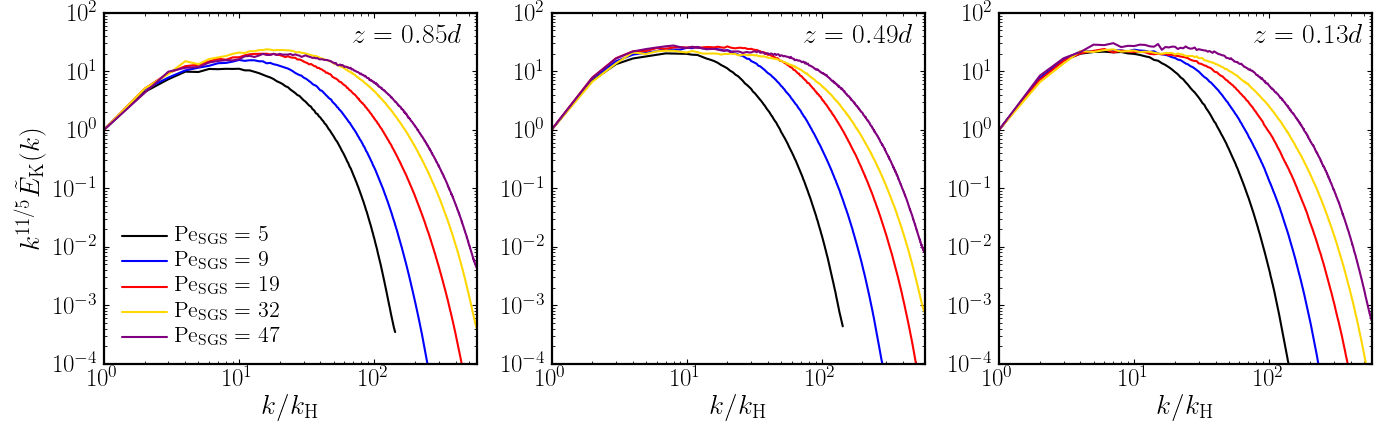}
\caption{Power spectra of the velocity compensated by $k^{11/5}$ from
  near the surface ($z=0.85d$, left panel), middle ($z=0.49d$,
  middle), and near the bottom of the CZ ($z=0.13d$, right) for
  $\PraSGS=0.1$ as a function of the SGS P\'eclet number as indicated
  by the legend. The spectra are normalized such that the
  contributions from $k/k_{\rm H}=1$ coincide.}
\label{plot_uu_spectra_PrSGS01}
\end{figure*}

\begin{figure*}
  \includegraphics[width=\textwidth]{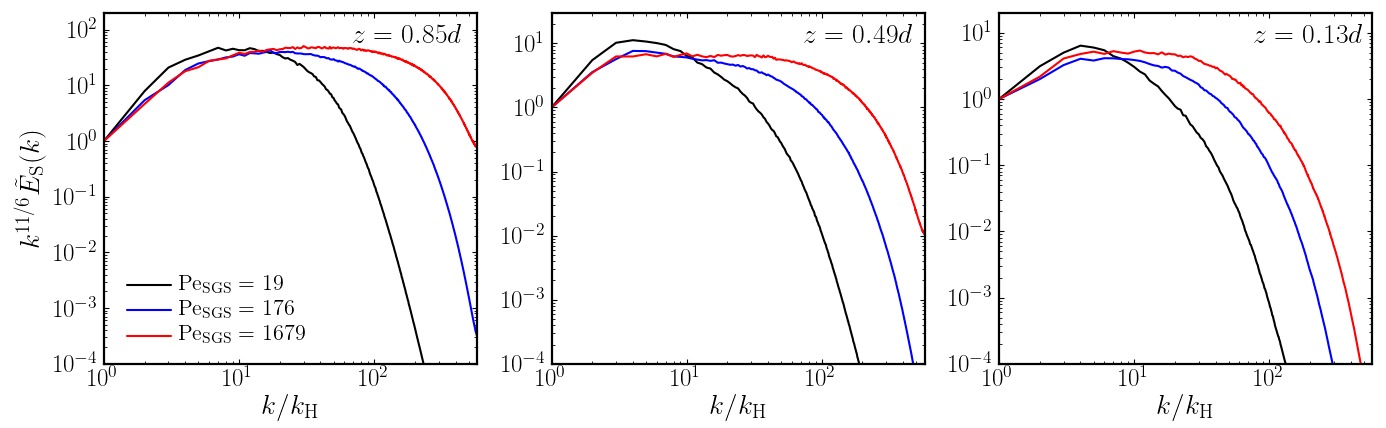}
\caption{Power spectra of entropy fluctuations $\tilde{E}_S(k)$
  compensated by $k^{11/6}$ from near the surface ($z=0.85d$, left
  panel), middle ($z=0.49d$, middle), and near the bottom of the CZ
  ($z=13d$, right) for runs C01 (black lines), C1 (blue), and C10
  (red). The spectra are normalized such that the contributions from
  $k/k_{\rm H}=1$ coincide.}
\label{fig:plot_ss_spectra_ad1152a3}
\end{figure*}

The current results indicate that increasing $\Rey$, and therefore
increasing $\Ra$, does not change the distribution of velocity power
in wavenumber space significantly. This is in apparent contradiction
with the results of \cite{FH16} who reported an increase of
small-scale flow amplitudes at the expense of large-scale power as the
Rayleigh number was increased. However, in their study the increase of
the Rayleigh number was associated with a significant change in the
fraction of the energy flux that is carried by convection because
their SGS entropy diffusion contributes to the mean energy flux, see
their Figure 6. It also appears that when the Rayleigh number is
sufficiently large, the decrease of the large-scale power ceases also
for \cite{FH16}; see their Figure 3. In the present study the
convective flux is not directly influenced by the SGS entropy
diffusion and thus the current results differ qualitatively from those
of \cite{FH16}.

\Figu{fig:plot_ss_spectra_ad1152a3} shows compensated power spectra of
specific entropy fluctuations from runs C01, C1, and C10. The spectra
are compensated by $k^{11/6}$ which appears to be compatible with the
highest $\Pe$ cases. In $\PraSGS=0.1$ there is no clear inertial range
due to the low P\'eclet number in run C01. In runs C1 and C10 where
the P\'eclet number is larger the entropy fluctuations show roughly a
$k^{-11/6}$ scaling at intermediate scales near the surface and at the
middle of the CZ (top and middle panels of
\Figa{fig:plot_ss_spectra_ad1152a3}). Near the base of the CZ, only
run C10 with the highest $\Pe$ shows signs of $k^{-11/6}$ spectra. The
observed scaling is steeper than those from the Kolmogorov-Obukhov and
Bolgiano-Obukhov models that predict $k^{-5/3}$ and $k^{-7/5}$,
respectively. Based on the current results, it appears that the
scaling of $\tilde{E}_S$ is becoming progressively shallower as $\Pe$
increases such that neither of the theoretical predictions can be
ruled out at the moment. However, simulations at even higher
resolutions are needed to confirm this.

\subsection{Convective energy transport}

\begin{figure}
  \includegraphics[width=\columnwidth]{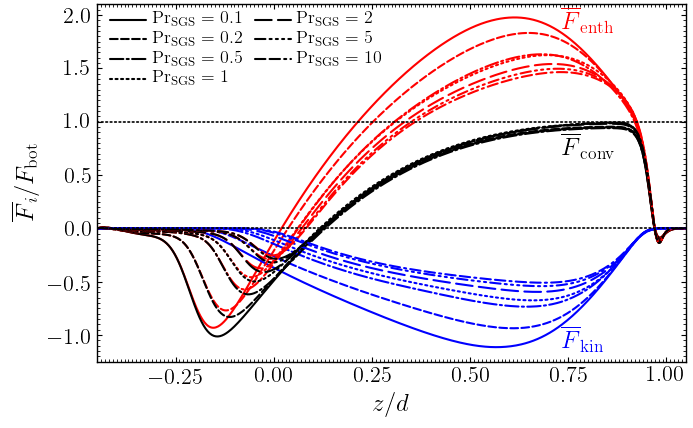}
\caption{Normalized enthalpy (red), kinetic energy (blue), and
  convected (black) fluxes for the SGS Prandtl numbers ranging from
  0.1 to 10 as indicated by the legend from the B set of runs with
  $\Rey = 79\ldots 93$.}
\label{fig:pconv_flux_ad576a3}
\end{figure}

\Figu{fig:pconv_flux_ad576a3} shows the, enthalpy, kinetic energy, and
total convected fluxes, Eqs.~(\ref{equ:Fenth}), (\ref{equ:Fkin}), and
(\ref{equ:convflux}), respectively, for the runs in the B set with
$\Rey = 79 \ldots 93$ and $576^3$ grid. Remarkably, $\mFconv$ is
practically identical within the CZ in all of the runs irrespective of
the SGS Prandtl number. This indicates that the radiative flux and
hence the thermal stratification and radiative conductivity are very
similar in all of the runs. The constituents of the convective flux,
however, show a very different behavior: the enthalpy flux increases
and the kinetic energy flux decreases monotonically with $\PraSGS$. In
particular, the kinetic energy flux more than doubles as $\PraSGS$
decreases from $10$ to $0.1$. For the lowest $\PraSGS$ the downward
kinetic energy flux exceeds $\Fbot$ near the middle of the CZ while
$\mFenth$ is almost twice $\Fbot$. In contrast to the CZ, the
convected flux in the overshoot layer shows much larger differences
between different Prandtl numbers.

\begin{figure}
  \includegraphics[width=\columnwidth]{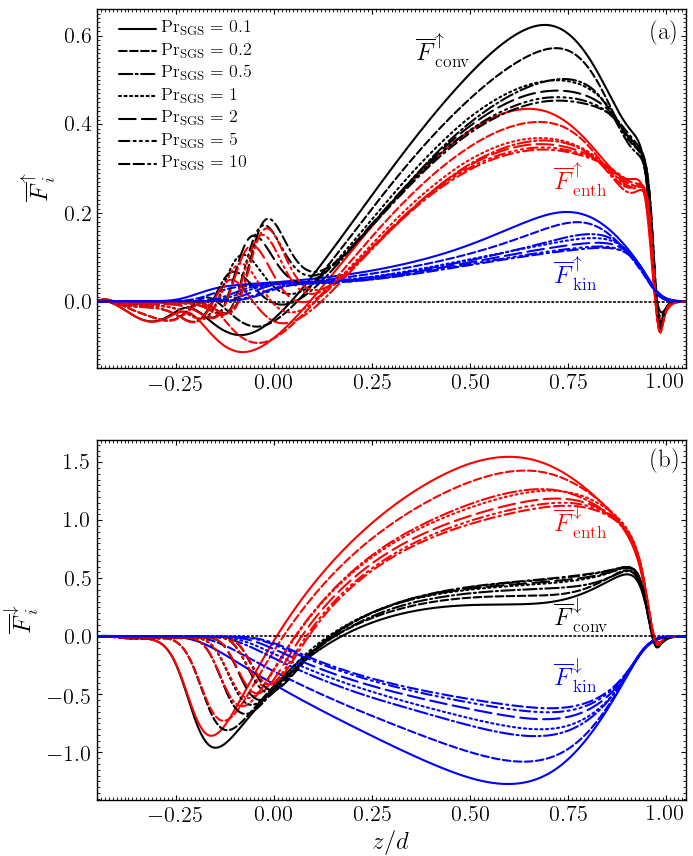}
\caption{Contributions from upflows \emph{(a)} and downflows
  \emph{(b)} on the horizontally averaged convected (black lines),
  enthalpy (red), and kinetic energy (blue) fluxes for the same runs
  as in \Figa{fig:pconv_flux_ad576a3}.}
\label{fig:pconv_flux_ud2_ad576a3}
\end{figure}

The contributions of up- and downflows to the enthalpy, kinetic energy
and total convected flux $\mFconv$ are shown in
\Figa{fig:pconv_flux_ud2_ad576a3}. The contributions of the upflows to
$\mFenth$ and $\mFkin$ increase monotonically as $\PraSGS$ decreases,
leading to a net increase of the convected flux
$\mFconv^\uparrow$. The magnitudes of $\mFenth^\downarrow$ and
$\mFkin^\downarrow$ also increase as the SGS Prandtl number decreases
but the net $\mFconv^\downarrow$ decreases because
$\mFenth^\downarrow$ and $\mFkin^\downarrow$ have opposite
signs. Remarkably, $\mFenth$ and $\mFkin$ are almost identical above
$z/d \approx 0.9$ irrespective of $\PraSGS$, suggesting that
near-surface physics are not the cause of the differences discussed
here. In total, the up- and downflows transport on average an equal
fraction of $\mFconv$ for SGS Prandtl number unity. The upflows are
clearly dominant for the lowest SGS Prandtl numbers such that for
$\PraSGS=0.1$ the upflows transport roughly two thirds of the
convected flux within the CZ. An opposite, albeit weaker, trend is
seen for $\PraSGS > 1$. The results for $\PraSGS\neq1$ are thus
qualitatively different from the case $\PraSGS = 1$, and the
difference increases toward large and small Prandtl numbers. The
current results are also in accordance with those of \cite{CBTMH91}
who found that in their more turbulent cases the contribution of the
downflows to $\Fconv$ was diminishing. The increase in the level of
turbulence in their cases was associated with a decreasing Prandtl
number $\sigma$. In their simulations the contribution of the
downflows to the convected flux is practically negligible at $\sigma =
0.1$, see their Fig.~14d.


\begin{figure}
  \includegraphics[width=\columnwidth]{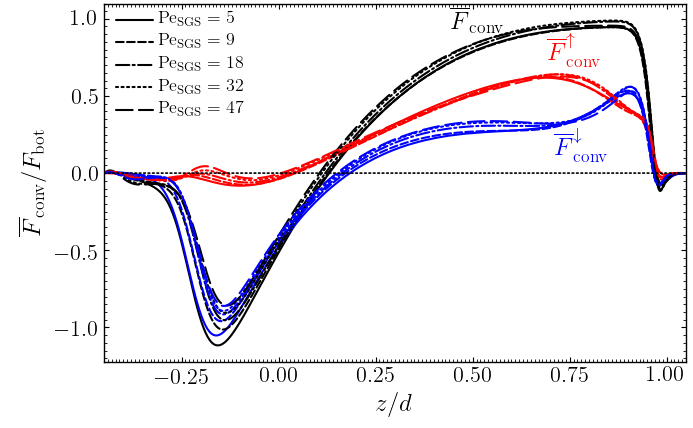}
\caption{Comparison of total convected flux (black) and the
  contributions of the upflows (red) and downflows (blue) between runs
  A01, B01, C01, C01b, and C01c with $\PraSGS=0.1$.}
\label{fig:pconv_flux_res}
\end{figure}

An important caveat of the current results is that changing $\PraSGS$
implies also that the P\'eclet number is changing. Thus it is
necessary to test whether the observed differences are really due to
the Prandtl number and not because of a P\'eclet number
dependence. Such a check is shown in \Figa{fig:pconv_flux_res} where
the convected flux from the five simulations with $\PraSGS=0.1$ (A01,
B01, C01, C01b, and C01c) are compared. The Reynolds and P\'eclet
numbers differ by an order of magnitude in this set of runs. The
differences are minor within the CZ whereas somewhat larger deviations
are seen in the depth of the overshoot region. Nevertheless, the
results are robust enough within the parameter range explored here
such that the conclusions drawn regarding the energy fluxes remain
valid. However, the current results should still be considered with
some caution as the P\'eclet numbers studied thus far are still modest
compared to realistic stellar conditions.

\begin{figure}
  \includegraphics[width=\columnwidth]{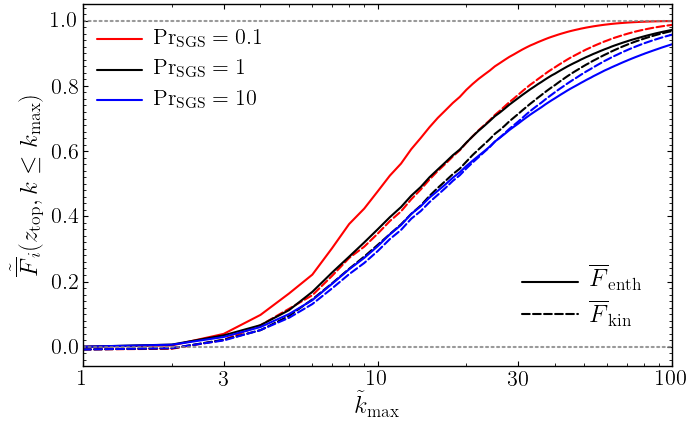}
  \caption{Spectrally decomposed normalized enthalpy
    ($\tilde{\overline{F}}_{\rm enth}$, solid lines) and kinetic
    energy ($\tilde{\overline{F}}_{\rm kin}$, dashed lines) fluxes
    from $z/d=0.84$ for runs C01 (red lines), C1 (black), and C10
    (blue).}
\label{fig:plot_fluxesk_ad1152a3_top}
\end{figure}

Although the statistics of velocity and entropy fluctuations do not
show drastic changes as a function of $\PraSGS$, yet the convective
energy transport is strongly affected. In the following the same
procedure as in \cite{2018ApJ...859..117N} is used to study the
dominant scales in the different contributions to the convective
flux. First, a low-pass filter is applied to the quantities that enter
the expressions of the fluxes such that wavenumbers $k<k_{\rm max}$
are retained. For the enthalpy (kinetic energy) flux this applies to
the fluctuating vertical momentum flux $(\rho u_z)'$ and temperature
fluctuation $T'$ (kinetic energy $\Ekin=\onehalf\rho\uuu^2$ and
vertical velocity $u_z$). Then the normalized, horizontally averaged,
enthalpy and kinetic energy fluxes up to $k_{\rm max}$ are computed
according to
\begin{eqnarray}
\tilde{\mean{F}}_{\rm enth}(k_{\rm max},z) &=& \nonumber \\ && \hspace{-1.2cm} \frac{1}{\Delta t} \int_{t_0}^{t_1} \left[ \frac{\mean{(\rho u_z)'(z,k_{\rm max},t) T'(z,k_{\rm max},t)}}{\mFenth(z,t)} \right] dt, \\ 
\tilde{\mean{F}}_{\rm kin}(k_{\rm max},z) &=& \nonumber \\ && \hspace{-1.2cm} \frac{1}{\Delta t} \int_{t_0}^{t_1} \left[ \frac{\mean{\Ekin(z,k_{\rm max},t) u_z(z,k_{\rm max},t)}}{\mFkin(z,t)} \right] dt.
\end{eqnarray}
Representative results for $\PraSGS = 0.1$, $1$, and $10$ are shown in
\Figa{fig:plot_fluxesk_ad1152a3_top} near the surface of the CZ. The
current results indicate that for $\PraSGS = 0.1$ the dominant
contribution of the enthalpy flux comes from larger scales than for
the kinetic energy flux. For $\PraSGS = 1$ the situation is
qualitatively unchanged but the difference in the dominant scales is
much smaller. On the other hand, for $\PraSGS = 10$ the behavior is
reversed although only at relatively large wavenumbers ($k/k_{\rm H}
\gtrsim 20$). These results suggest that while the dominant velocity
scale is quite insensitive to changes of the Prandtl number, a
stronger dependence exists in the convective energy transport.

The current results thus suggest that convection becomes less
efficient when the Prandtl number is decreased such that a larger
vertical velocities are required to carry the same flux. This has some
parallel in Rayleigh-B\'enard convection where the Nusselt number
decreases strongly for a given Rayleigh number when $\Pra$ is
decreased \citep[][and references therein]{2020RvMP...92d1001S}. This
means that low Prandtl number convection is very turbulent but at the
same time inefficient in transporting heat. The analogy to the
Reyleigh-B\'enard case is, however, not complete as in the current
simulations the ratio of convective to radiative flux is almost
unchanged when the SGS Prandtl number changes. Therefore the effects
of the Prandtl number are necessarily much more subtle in the present
cases and there is no straightforward way to connect, for example, the
average rms-velocity and the Rayleigh number.

\subsection{Velocity, temperature, and density fluctuations}

\begin{figure}
  \includegraphics[width=\columnwidth]{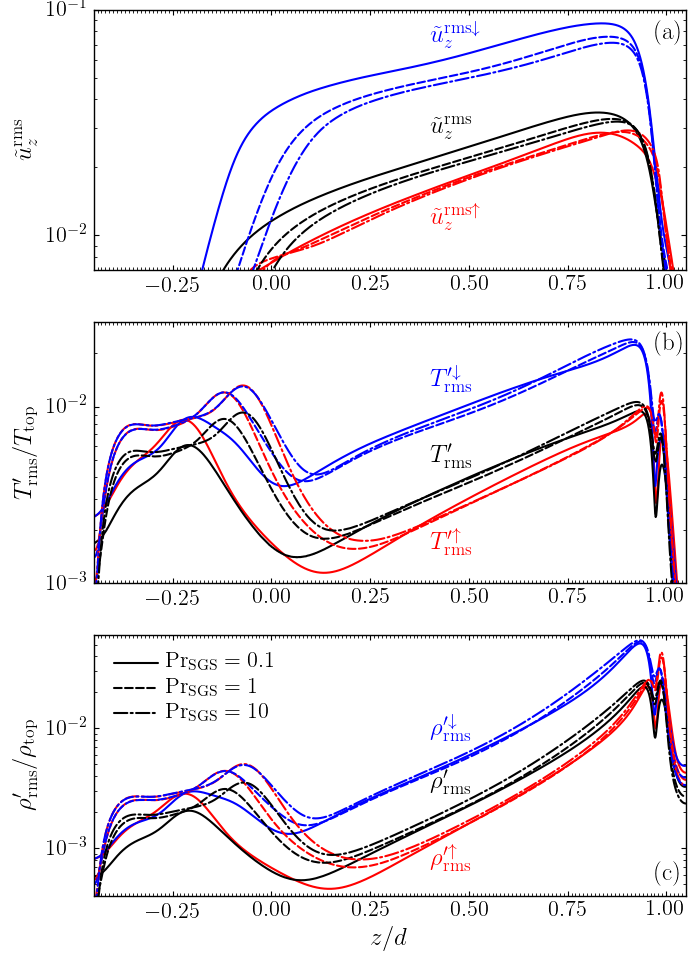}
\caption{Horizontally averaged total rms vertical velocity (black
  lines) and the contributions from upflows (red) and downflows (blue)
  \emph{(a)}. Corresponding rms temperature \emph{(b)} and density
  \emph{(c)} fluctuations normalized by $T_{\rm top}$ and $\rho_{\rm
    top}=\rho_0$, respectively. Data for runs C01, C1, and C10 with
  $\PraSGS=0.1$, $1$, and $10$ are shown as indicated by the legend.}
\label{fig:pucsf_ad1152a3}
\end{figure}

The fact that the magnitudes of enthalpy and kinetic energy fluxes
both increase as the SGS Prandtl number decreases while the total
convected energy flux remains constant, suggests a systematic change
in the velocity and temperature fluctuations or their correlation as a
function of $\PraSGS$ \citep[for the latter, see][]{BCNS05}. The
rms-fluctuations of $u_z$ for the total velocity, upflows, and
downflows are shown in \Figa{fig:pucsf_ad1152a3}(a) for runs C01, C1,
and C10. The rms vertical velocity in the bulk of the CZ increases
monotonically for both up- and downflows as $\PraSGS$ decreases. The
increase is particularly pronounced for the downflows, $\uzrmsdo$,
which is also the main contributor in the increase of $\uzrms$. This
reflects the decreasing filling factor of downflows with $\PraSGS$
(see \Figa{fig:fillingfactor}(a)) such that the downflows need to be
faster due to mass conservation.

The temperature fluctuations, shown in \Figa{fig:pucsf_ad1152a3}(b),
show a somewhat more complex behavior: near the surface $T'_{\rm rms}$
increases with $\PraSGS$ whereas in deeper parts the trend is
reversed. This can be understood such that near the surface the
temperature fluctuations are mostly in small-scale structures such as
granules and intergranular lanes which are subject to stronger
diffusion for smaller $\PraSGS$, resulting in smaller $T_{\rm rms}'$
on average there. In the bulk of the CZ, $T_{\rm rms}'$ is the largest
for the smallest $\PraSGS$ for both up- and downflows. This can partly
explain the dominance of upflows in the energy transport for
$\PraSGS=0.1$. For $\PraSGS=1$ and $10$, $T'_{\rm rms}$ is always
larger in the latter, which is due to the lower thermal
diffusivity. Finally, \Figa{fig:pucsf_ad1152a3}(c) shows the depth
dependence of the rms density fluctuations. Here a monotonic trend is
seen such that $\rho'_{\rm rms}$ decreases with $\PraSGS$.

\begin{figure}
  \includegraphics[width=\columnwidth]{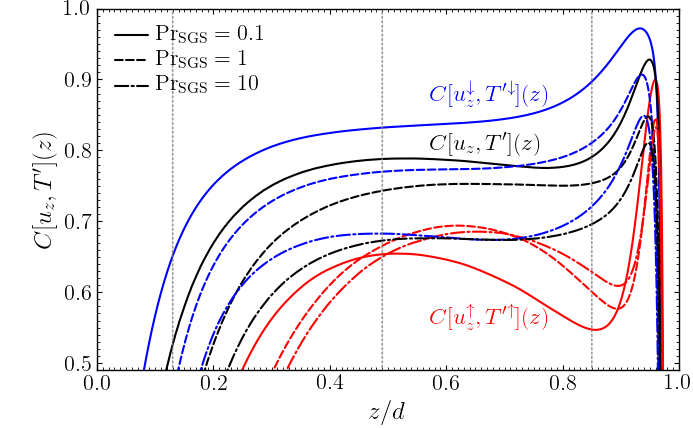}
\caption{Correlation coefficient $C[u_z,T']$ as a function of depth
  from run C01 ($\PraSGS=0.1$, solid lines), C1 ($\PraSGS=1$, dashed)
  and C10 ($\PraSGS=10$, dash-dotted) for the downflows (blue),
  upflows (red), and totals (black).}
\label{pcorruT_ad1152a3}
\end{figure}

The correlation coefficient of vertical velocity and temperature
fluctuations is given by
\begin{eqnarray}
C[u_z,T'] = \frac{\mean{u_z T'}}{\uzrms T'_{\rm rms}} \approx \frac{\mFenth}{\cP \mean{\rho} \uzrms T'_{\rm rms}},
\end{eqnarray}
where $\mFenth=\cP \mean{(\rho u_z)'T'}\approx\cP \mean{\rho}
\mean{u_z'T'}$. The correlation coefficients $C[u_z,T']$ for the total
flow, and up- and dowflows for the same runs as in
\Figa{fig:pucsf_ad1152a3} are shown in \Figa{pcorruT_ad1152a3}. The
overall correlation coefficient $C[u_z,T']$ decreases monotonically as
$\PraSGS$ increases in agreement with \cite{1993A&A...279..107S}. This
trend is dominated by the downflows where
$C[u_z^\downarrow,T'^\downarrow]$ increases with decreasing $\PraSGS$
while a much weaker and opposite trend is observed for the upflows.
The increasing correlation $C[u_z^\downarrow,T'^\downarrow]$ with
decreasing $\PraSGS$ explains the simultaneous strong increase of the
magnitude of $\Fenth^\downarrow$. At the same time the correlation in
the upflows is much less affected by the SGS Prandtl number and the
overall increase of vertical velocity and temperature fluctuations for
low $\PraSGS$ (see, \Figa{fig:pucsf_ad1152a3}) explains the increased
$\Fenth^\uparrow$ in that regime.

\begin{figure}
  \includegraphics[width=\columnwidth]{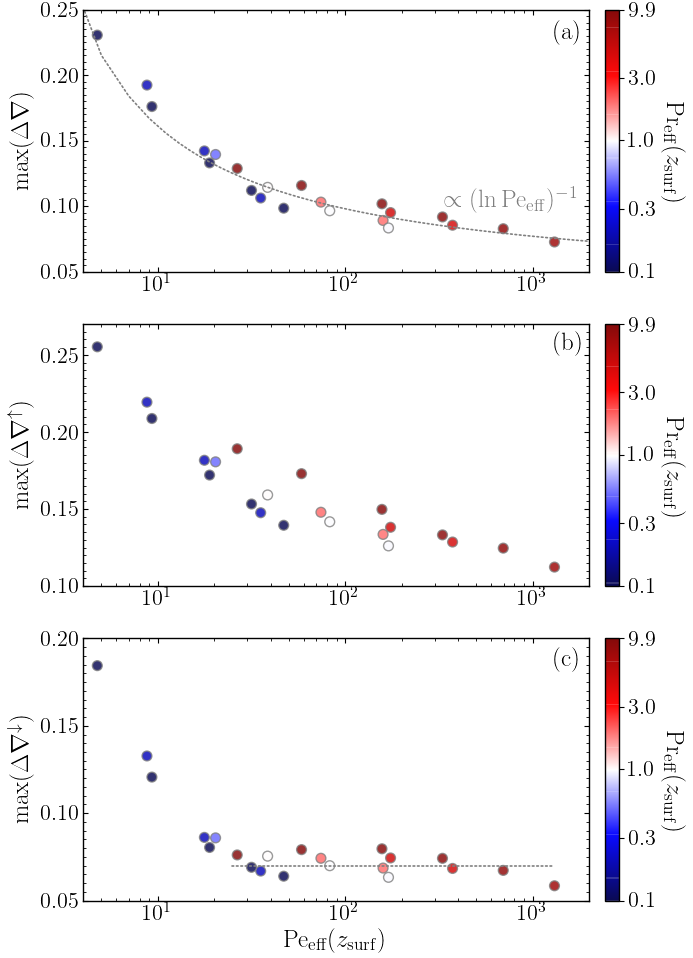}
\caption{Maximum of the superadiabatic temperature gradient
  $\Delta\nabla = \nabla-\nabad$ as a function of the effective
  P\'eclet number at $z_{\rm surf} = 0.85d$ for the total \emph{(a)},
  in upflows \emph{(b)}, and in downflows \emph{(c)}. The colour
  coding indicates $\Pra_{\rm eff}(z_{\rm surf})$.}
\label{fig:psamax}
\end{figure}

\subsection{Driving of convection}

The dependence of the convective flows on the Prandtl number raises
the question of their driving. A primary candidate is the entropy
gradient at the surface which has indeed reported to be Prandtl number
dependent \citep[e.g.][]{BCNS05}. A diagnostic of this is the maximum
value of the superadiabatic temperature gradient
\begin{eqnarray}
\Delta \nabla = \nabla - \nabad = -\frac{H_p}{\cP} \frac{ds}{dz},
\end{eqnarray}
where $\nabla = \pd \ln T/\pd \ln p$ and $\nabad = 1 - 1/\gamma$ are
the logarithmic and adiabatic temperature gradients,
respectively. \Figu{fig:psamax} shows the maximum value of $\Delta
\nabla$ near the surface for the total entropy as well as for the
upflow and downflow regions separately for all of the current
runs. The results for $\mbox{max}(\Delta\nabla)$ show a decreasing
trend as a function of $\Pe_{\rm eff}$ and only a much weaker
dependence on $\Praeff$. The data perhaps suggests a $(\ln \Pe_{\rm
  eff})^{-1}$ dependence. A similar decreasing trend is seen in the
upflow regions although the scatter in the data is stronger especially
in the intermediate range of $\Pe_{\rm eff}$. On the other hand, the
data for the downflows perhaps suggests a plateau for $\Pe_{\rm
  eff}\gtrsim 30$. However, these tentative dependences are rather
uncertain due to the limited data available. Nevertheless, it appears
that for approximately the same P\'eclet number the entropy gradient
increases with the Prandtl number. This is opposite to the trend in
the strength of downflows and thus the surface entropy gradient cannot
be the dominant contribution in driving the downflows.

\begin{figure}
  \includegraphics[width=\columnwidth]{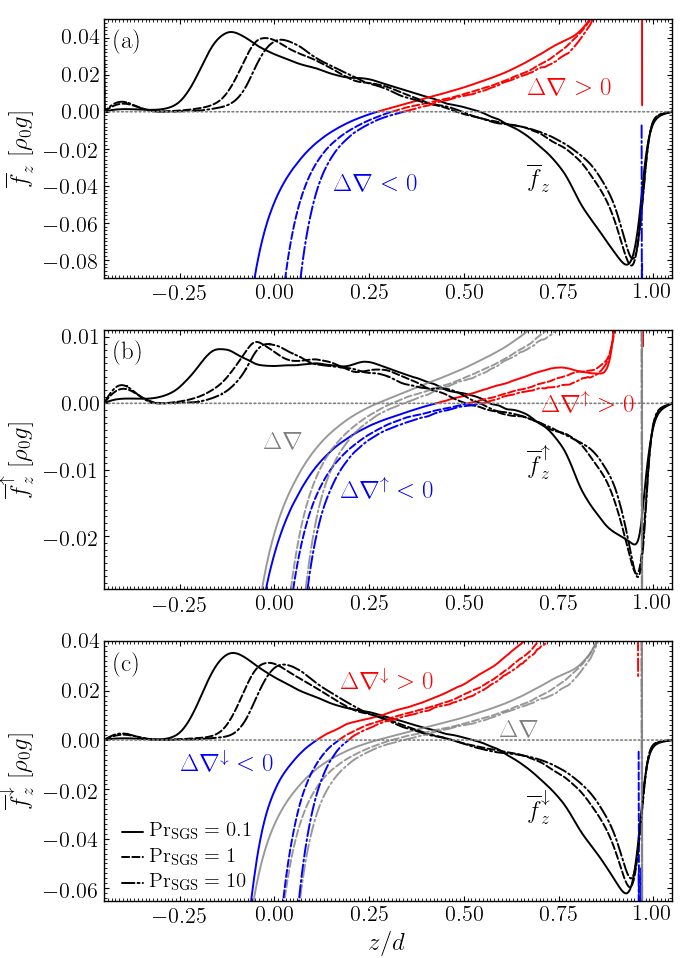}
  \caption{Horizontally averaged forces for the total flow \emph{(a)},
    upflows \emph{(b)} and downflows \emph{(c)} for runs C01
    ($\PraSGS=0.1$, solid lines), C1 ($\PraSGS=1$, dashed), and C10
    ($\PraSGS=10$ dash-dotted). The blue and red lines show the
    corresponding superadiabatic temperature gradients with blue
    denoting negative and red positive values. The grey lines in
    panels \emph{(b)} and \emph{(c)} indicate $\Delta \nabla$ for
    reference.}
\label{plot_xyaver_forces_ad1152a3}
\end{figure}

Next we turn to the force balance for vertical flows. The horizontally
averaged force density is given by
\begin{eqnarray}
\mean{f}_z = \mean{\rho {\cal D} u_z/{\cal D}t},
\end{eqnarray}
where ${\cal D}/{\cal D}t = \pd/\pd t - u_z \pd_z$ is the vertical
advective time derivative. Representative results for the total force
are shown in \Figa{plot_xyaver_forces_ad1152a3}(a) for runs C01, C1,
and C10. The runs with SGS Prandtl numbers 1 and 10 show a small
difference near the surface in that the maximum total force is
slightly larger for the lower Prandtl number. On the other hand,
$\mean{f}_z$ for $\PraSGS = 0.1$ shows a markedly wider negative
region in the upper part of the CZ between $0.65\lesssim z/d \lesssim
0.9$. The force on the upflows is shown in
\Figa{plot_xyaver_forces_ad1152a3}(b). The differences between the two
higher SGS Prandtl numbers are very small whereas the $\PraSGS = 0.1$
case again shows a clear difference in the near-surface layers such
that the region where upflows are decelerated is significantly
wider. Remarkably, $\mean{f}_z^\uparrow$ is in almost perfect
anticorrelation with the superadiabatic temperature gradient in the
upflows ($\Delta \nabla^\uparrow$) such that upflows are decelerated
(accelerated) in unstably (stably) stratified regions. This suggests
that the upflows are not directly driven by the convective instability
itself but rather by pressure forces induced by the deeply penetrating
downflows \citep[see
  also][]{2017PhRvE..96c3104K,2017ApJ...845L..23K}. Finally,
\Figa{plot_xyaver_forces_ad1152a3}(c) shows the forces for the
downflows. The downward force on the downflows is larger near the
surface for $\PraSGS=0.1$ than in the cases with $\PraSGS=1$ and $10$
which explains the stronger vertical velocities for low SGS Prandtl
numbers. The force on the downflows is almost in anticorrelation with
the overall superadiabatic temperature gradient; compare the grey and
black lines in \Figa{plot_xyaver_forces_ad1152a3}(c).\footnote{In
  earlier studies of non-rotating hydrodynamic convection
  \citep{2017ApJ...845L..23K,2019A&A...631A.122K} the total force on
  the downflows was found to adhere closely to $\Delta \nabla$. These
  analyses, however, contain an error due to which the contribution
  from the viscous force is underestimated by a factor roughly between
  two and 30 depending on the depth in the CZ. Thus the agreement
  between $\Delta \nabla$ and $\mean{f}_z^\uparrow$ in these earlier
  studies is poorer than reported and comparable to that presented in
  the current study.} The correlation with $\Delta \nabla^\downarrow$
is clearly poorer.

\Figa{fig:pucsf_ad1152a3}(c) showed that the density fluctuations are
weakly decreasing with $\PraSGS$ such that the increased acceleration
on the downflows cannot be explained by arguing that the matter in the
downflows is cooler and heavier in the low $\PraSGS$ cases in
comparison to the higher $\PraSGS$ cases. On the other hand,
\Figa{pcorruT_ad1152a3} shows that the correlation between vertical
velocity and temperature fluctuations increases with decreasing
$\PraSGS$ which is the decisive factor in the enhanced acceleration of
downflows. A similar enhancement of correlation is likely to carry
over to other thermodynamic quantities as well. It is not clear what
the exact mechanims is but it seems plausible to assume that a process
similar to the positive feedback loop between vertical flows and the
buoyancy force in the zero Prandtl number limit in Rayleigh-B\'enard
convection \citep{1962JGR....67.3063S} is present also in the
compressible case.

\section{Conclusions}

Convective energy transport and flow statistics are sensitive to the
SGS Prandtl number in the parameter range currently accessible to
numerical simulations. The most striking effect is the decrease of net
energy transport due to downflows with decreasing Prandtl number. This
happens because the oppositely signed enthalpy and kinetic energy
fluxes both increase in the dowflows, leading to increased
cancellation \citep{CBTMH91}. Another effect of a decreasing Prandtl
number is the increase of the overall velocity which is dominantly due
to the increase of the downflow strength. The stronger downflows also
lead to much stronger overshooting at the base of the CZ for lower
Prandtl number.

On the other hand, the effect of the Prandtl number are very subtle in
the statistics of the velocity field. The clearest systematic effect
is that the filling factor of downflows decreases monotonically with
decreasing SGS Prandtl number. However, the power spectra of velocity
show very small differences apart from the deep layers where the
downflow dominance in the low-Prandtl number regime becomes more
prominent. The current results do not indicate a decrease of the large
scale velocity power neither with decreasing Prandtl nor with
increasing Rayleigh numbers. Such decrease has been suggested to be at
least a partial solution to the convective conundrum, or the too high
large-scale power in simulated flows \citep{FH16}. In the current
simulations the large-scale power is almost unaffected most likely
because the SGS entropy diffusion does not contribute to the mean
energy flux unlike in the simulations of \cite{FH16}. Notably,
however, the dominant spatial scales of enthalpy and kinetic energy
fluxes vary systematically with $\PraSGS$ which is likely to hold the
key to understanding the changing convective dynamics.

The implications of the current results of solar and stellar
convection are difficult to assess because of the greatly differing
parameter regimes of the simulations compared to stellar CZs. Another
aspect is that the current simulations are very likely not in an
asymptotic regime such that the results show a dependence on the
Reynolds and P\'eclet numbers. Nevertheless, even with these
reservations, it appears likely that convection in the Sun is quite
different from that obtained from simulations in which $\Pra \approx
1$. In particular, the overshooting depth can be substantially
underestimated by the current simulations. It is also unclear how the
Prandtl number effects manifest themselves in angular momentum
transport which has thus far been only discussed in the $\Pra \gtrsim
1$ regime \citep{2018PhFl...30d6602K}.

The current simulations use SGS diffusion for the entropy fluctuations
which is solely because of numerical convenience. The use of SGS
diffusion has been criticized because it has no physical counterpart
and as a possible cause for unrealistically strong energy fluxes in
the downflows. Addressing these question with simulations where the
SGS diffusion is absent will be presented elsewhere.

\begin{acknowledgements}
  I thank Axel Brandenburg and Nishant Singh for their comments on the
  manuscript. I acknowledge the hospitality of Nordita during the
  program `The Shifting Paradigm of Stellar Convection: From Mixing
  Length Concepts to Realistic Turbulence Modelling'. The simulations
  were made within the Gauss Center for Supercomputing project
  ``Cracking the Convective Conundrum'' in the Leibniz Supercomputing
  Centre's SuperMUC--NG supercomputer in Garching, Germany. This work
  was supported by the Deutsche Forschungsgemeinschaft Heisenberg
  programme (grant No.\ KA 4825/2-1).
\end{acknowledgements}

\bibliographystyle{aa}
\bibliography{paper}

\end{document}

%% file: macros.tex
\newcommand{\uuu}{{\bm u}}

\newcommand{\xxx}{{\bm x}}

\newcommand{\FFF}{{\bm F}}

\newcommand{\SSt}{\bm{\mathsf{S}}}
\newcommand{\SStij}{\mathsf{S}_{ij}}
\newcommand{\Eq}[1]{Eq.~(\ref{#1})}

\newcommand{\Eqsa}[2]{Eqs.~(\ref{#1}) and (\ref{#2})} 
\newcommand{\Equsa}[2]{Eqs.~(\ref{#1}) to (\ref{#2})}
\newcommand{\EQ}{\begin{equation}}
\newcommand{\EN}{\end{equation}}
\newcommand{\EQA}{\begin{eqnarray}}
\newcommand{\ENA}{\end{eqnarray}}

\newcommand{\pd}{\partial}

\newcommand{\mean}[1]{\overline{#1}}

\newcommand{\cP}{c_{\rm P}}
\newcommand{\cV}{c_{\rm V}}

\newcommand{\urms}{u_{\rm rms}}

\newcommand{\uzrms}{u_z^{\rm rms}}
\newcommand{\uzrmsdo}{u_z^{{\rm rms}\downarrow}}

\newcommand{\chiSGS}{\chi_{\rm SGS}}

\newcommand{\chiSGSo}{\chi_{\rm SGS}^{(1)}}

\newcommand{\Pe}{{\rm Pe}}
\newcommand{\Pra}{{\rm Pr}}
\newcommand{\Praeff}{{\rm Pr}_{\rm eff}}
\newcommand{\PraSGS}{{\rm Pr}_{\rm SGS}}

\newcommand{\Ra}{{\rm Ra}}

\newcommand{\Rey}{{\rm Re}}

{}

\newcommand{\calR}{{\cal R}}

\newcommand{\Fenth}{\mean{F}_{\rm enth}}

\newcommand{\Fconv}{F_{\rm conv}}
\newcommand{\Frad}{F_{\rm rad}}

\newcommand{\Fbot}{F_{\rm bot}}

\newcommand{\Fn}{\mathscr{F}_{\rm n}}
\newcommand{\mFenth}{\mean{F}_{\rm enth}}
\newcommand{\mFconv}{\mean{F}_{\rm conv}}
\newcommand{\mFrad}{\mean{F}_{\rm rad}}
\newcommand{\mFkin}{\mean{F}_{\rm kin}}
\newcommand{\mFvisc}{\mean{F}_{\rm visc}}

\newcommand{\mFcool}{\mean{F}_{\rm cool}}

\newcommand{\Ekin}{E_{\rm kin}}

\newcommand{\nabad}{\nabla_{\rm ad}}

\newcommand{\tbot}{{\rm bot}}

\newcommand{\zbot}{z_{\rm bot}}
\newcommand{\zbz}{z_{\rm BZ}}
\newcommand{\zcz}{z_{\rm CZ}}
\newcommand{\zos}{z_{\rm OS}}
\def\onethird{{\textstyle{1\over3}}}

\def\onehalf{{\textstyle{1\over2}}}

\newcommand{\Figa}[1]{Fig.~\ref{#1}}

\newcommand{\Figu}[1]{Figure~\ref{#1}}

\newcommand{\Sec}[1]{Section~\ref{#1}} 
\newcommand{\Table}[1]{Table~\ref{#1}}


%% file: paper.bbl
\begin{thebibliography}{56}
\expandafter\ifx\csname natexlab\endcsname\relax\def\natexlab#1{#1}\fi

\bibitem[{{Augustson} {et~al.}(2019){Augustson}, {Brun}, \&
  {Toomre}}]{2019ApJ...876...83A}
{Augustson}, K.~C., {Brun}, A.~S., \& {Toomre}, J. 2019, \apj, 876, 83

\bibitem[{{Barekat} \& {Brandenburg}(2014)}]{BB14}
{Barekat}, A. \& {Brandenburg}, A. 2014, \aap, 571, A68

\bibitem[{{Bekki} {et~al.}(2017){Bekki}, {Hotta}, \&
  {Yokoyama}}]{2017ApJ...851...74B}
{Bekki}, Y., {Hotta}, H., \& {Yokoyama}, T. 2017, \apj, 851, 74

\bibitem[{{Bolgiano}(1959)}]{1959JGR....64.2226B}
{Bolgiano}, R., J. 1959, \jgr, 64, 2226

\bibitem[{{Brandenburg}(1992)}]{1992PhRvL..69..605B}
{Brandenburg}, A. 1992, \prl, 69, 605

\bibitem[{{Brandenburg}(2003)}]{B03}
{Brandenburg}, A. 2003, {Computational aspects of astrophysical MHD and
  turbulence}, ed. A.~{Ferriz-Mas} \& M.~{N{\'u}{\~n}ez} (London: Taylor and
  Francis), 269

\bibitem[{{Brandenburg}(2016)}]{Br16}
{Brandenburg}, A. 2016, \apj, 832, 6

\bibitem[{{Brandenburg} {et~al.}(2005){Brandenburg}, {Chan}, {Nordlund}, \&
  {Stein}}]{BCNS05}
{Brandenburg}, A., {Chan}, K.~L., {Nordlund}, {\AA}., \& {Stein}, R.~F. 2005,
  AN, 326, 681

\bibitem[{{Brandenburg} {et~al.}(1996){Brandenburg}, {Jennings}, {Nordlund},
  {Rieutord}, {Stein}, \& {Tuominen}}]{BJNRST96}
{Brandenburg}, A., {Jennings}, R.~L., {Nordlund}, {\AA}., {et~al.} 1996, J.
  Fluid Mech., 306, 325

\bibitem[{{Brandenburg} {et~al.}(2000){Brandenburg}, {Nordlund}, \&
  {Stein}}]{2000gac..conf...85B}
{Brandenburg}, A., {Nordlund}, A., \& {Stein}, R.~F. 2000, in Geophysical and
  Astrophysical Convection, Contributions from a workshop sponsored by the
  Geophysical Turbulence Program at the National Center for Atmospheric
  Research, October, 1995. Edited by Peter A. Fox and Robert M. Kerr. Published
  by Gordon and Breach Science Publishers, The Netherlands, 2000, p. 85-105,
  ed. P.~A. {Fox} \& R.~M. {Kerr}, 85--105

\bibitem[{{Breuer} {et~al.}(2004){Breuer}, {Wessling}, {Schmalzl}, \&
  {Hansen}}]{2004PhRvE..69b6302B}
{Breuer}, M., {Wessling}, S., {Schmalzl}, J., \& {Hansen}, U. 2004, \pre, 69,
  026302

\bibitem[{{Calzavarini} {et~al.}(2002){Calzavarini}, {Toschi}, \&
  {Tripiccione}}]{2002PhRvE..66a6304C}
{Calzavarini}, E., {Toschi}, F., \& {Tripiccione}, R. 2002, \pre, 66, 016304

\bibitem[{{Cattaneo} {et~al.}(1991){Cattaneo}, {Brummell}, {Toomre},
  {Malagoli}, \& {Hurlburt}}]{CBTMH91}
{Cattaneo}, F., {Brummell}, N.~H., {Toomre}, J., {Malagoli}, A., \& {Hurlburt},
  N.~E. 1991, \apj, 370, 282

\bibitem[{{Chan} \& {Gigas}(1992)}]{CG92}
{Chan}, K.~L. \& {Gigas}, D. 1992, \apjl, 389, L87

\bibitem[{{Chan} \& {Sofia}(1986)}]{CS86}
{Chan}, K.~L. \& {Sofia}, S. 1986, \apj, 307, 222

\bibitem[{{Cossette} \& {Rast}(2016)}]{CR16}
{Cossette}, J.-F. \& {Rast}, M.~P. 2016, \apjl, 829, L17

\bibitem[{{Deardorff}(1961)}]{1961JAtS...18..540D}
{Deardorff}, J.~W. 1961, J. Atmosph. Sci., 18, 540

\bibitem[{{Deardorff}(1966)}]{De66}
{Deardorff}, J.~W. 1966, J. Atmosph. Sci., 23, 503

\bibitem[{{Dobler} {et~al.}(2006){Dobler}, {Stix}, \& {Brandenburg}}]{DSB06}
{Dobler}, W., {Stix}, M., \& {Brandenburg}, A. 2006, \apj, 638, 336

\bibitem[{{Fan} \& {Fang}(2014)}]{FF14}
{Fan}, Y. \& {Fang}, F. 2014, \apj, 789, 35

\bibitem[{{Featherstone} \& {Hindman}(2016)}]{FH16}
{Featherstone}, N.~A. \& {Hindman}, B.~W. 2016, \apj, 818, 32

\bibitem[{{Greer} {et~al.}(2015){Greer}, {Hindman}, {Featherstone}, \&
  {Toomre}}]{GHFT15}
{Greer}, B.~J., {Hindman}, B.~W., {Featherstone}, N.~A., \& {Toomre}, J. 2015,
  \apjl, 803, L17

\bibitem[{{Hanasoge} {et~al.}(2016){Hanasoge}, {Gizon}, \&
  {Sreenivasan}}]{2016AnRFM..48..191H}
{Hanasoge}, S., {Gizon}, L., \& {Sreenivasan}, K.~R. 2016, Annual Review of
  Fluid Mechanics, 48, 191

\bibitem[{{Hanasoge} {et~al.}(2012){Hanasoge}, {Duvall}, \&
  {Sreenivasan}}]{HDS12}
{Hanasoge}, S.~M., {Duvall}, T.~L., \& {Sreenivasan}, K.~R. 2012, Proc. Natl.
  Acad. Sci., 109, 11928

\bibitem[{{Hotta}(2017)}]{2017ApJ...843...52H}
{Hotta}, H. 2017, \apj, 843, 52

\bibitem[{{Hotta} {et~al.}(2019){Hotta}, {Iijima}, \&
  {Kusano}}]{2019SciA....5.2307H}
{Hotta}, H., {Iijima}, H., \& {Kusano}, K. 2019, Science Advances, 5, 2307

\bibitem[{{Hotta} {et~al.}(2015){Hotta}, {Rempel}, \& {Yokoyama}}]{HRY15b}
{Hotta}, H., {Rempel}, M., \& {Yokoyama}, T. 2015, \apj, 803, 42

\bibitem[{{K{\"a}pyl{\"a}}(2011)}]{K11}
{K{\"a}pyl{\"a}}, P.~J. 2011, Astron. Nachr., 332, 43

\bibitem[{{K{\"a}pyl{\"a}}(2019{\natexlab{a}})}]{2019AN....340..744K}
{K{\"a}pyl{\"a}}, P.~J. 2019{\natexlab{a}}, Astronomische Nachrichten, 340, 744

\bibitem[{{K{\"a}pyl{\"a}}(2019{\natexlab{b}})}]{2019A&A...631A.122K}
{K{\"a}pyl{\"a}}, P.~J. 2019{\natexlab{b}}, \aap, 631, A122

\bibitem[{{K{\"a}pyl{\"a}}(2021)}]{2020arXiv201201259K}
{K{\"a}pyl{\"a}}, P.~J. 2021, arXiv:2012.01259

\bibitem[{{K{\"a}pyl{\"a}} {et~al.}(2014){K{\"a}pyl{\"a}}, {K{\"a}pyl{\"a}}, \&
  {Brandenburg}}]{KKB14}
{K{\"a}pyl{\"a}}, P.~J., {K{\"a}pyl{\"a}}, M.~J., \& {Brandenburg}, A. 2014,
  \aap, 570, A43

\bibitem[{{K{\"a}pyl{\"a}} {et~al.}(2017){K{\"a}pyl{\"a}}, {Rheinhardt},
  {Brandenburg}, {Arlt}, {K{\"a}pyl{\"a}}, {Lagg}, {Olspert}, \&
  {Warnecke}}]{2017ApJ...845L..23K}
{K{\"a}pyl{\"a}}, P.~J., {Rheinhardt}, M., {Brandenburg}, A., {et~al.} 2017,
  \apjl, 845, L23

\bibitem[{{K{\"a}pyl{\"a}} {et~al.}(2019){K{\"a}pyl{\"a}}, {Viviani},
  {K{\"a}pyl{\"a}}, {Brandenburg}, \& {Spada}}]{2019GApFD.113..149K}
{K{\"a}pyl{\"a}}, P.~J., {Viviani}, M., {K{\"a}pyl{\"a}}, M.~J., {Brandenburg},
  A., \& {Spada}, F. 2019, Geophysical and Astrophysical Fluid Dynamics, 113,
  149

\bibitem[{{Karak} {et~al.}(2018){Karak}, {Miesch}, \&
  {Bekki}}]{2018PhFl...30d6602K}
{Karak}, B.~B., {Miesch}, M., \& {Bekki}, Y. 2018, Physics of Fluids, 30,
  046602

\bibitem[{{Korre} {et~al.}(2017){Korre}, {Brummell}, \&
  {Garaud}}]{2017PhRvE..96c3104K}
{Korre}, L., {Brummell}, N., \& {Garaud}, P. 2017, \pre, 96, 033104

\bibitem[{{Kupka} \& {Muthsam}(2017)}]{2017LRCA....3....1K}
{Kupka}, F. \& {Muthsam}, H.~J. 2017, Liv. Rev. Comp. Astrophys., 3, 1

\bibitem[{{Miesch} {et~al.}(2008){Miesch}, {Brun}, {DeRosa}, \&
  {Toomre}}]{2008ApJ...673..557M}
{Miesch}, M.~S., {Brun}, A.~S., {DeRosa}, M.~L., \& {Toomre}, J. 2008, \apj,
  673, 557

\bibitem[{{Nelson} {et~al.}(2018){Nelson}, {Featherstone}, {Miesch}, \&
  {Toomre}}]{2018ApJ...859..117N}
{Nelson}, N.~J., {Featherstone}, N.~A., {Miesch}, M.~S., \& {Toomre}, J. 2018,
  \apj, 859, 117

\bibitem[{{Obukhov}(1959)}]{1959DoSSR..125..1246O}
{Obukhov}, A.~M. 1959, Akademiia Nauk SSSR Doklady, 125, 1246

\bibitem[{{O'Mara} {et~al.}(2016){O'Mara}, {Miesch}, {Featherstone}, \&
  {Augustson}}]{2016AdSpR..58.1475O}
{O'Mara}, B., {Miesch}, M.~S., {Featherstone}, N.~A., \& {Augustson}, K.~C.
  2016, Adv. Space Res., 58, 1475

\bibitem[{{Orvedahl} {et~al.}(2018){Orvedahl}, {Calkins}, {Featherstone}, \&
  {Hindman}}]{2018ApJ...856...13O}
{Orvedahl}, R.~J., {Calkins}, M.~A., {Featherstone}, N.~A., \& {Hindman}, B.~W.
  2018, \apj, 856, 13

\bibitem[{{Ossendrijver}(2003)}]{O03}
{Ossendrijver}, M. 2003, \aapr, 11, 287

\bibitem[{{Pencil Code Collaboration} {et~al.}(2021){Pencil Code
  Collaboration}, {Brandenburg}, {Johansen}, {Bourdin}, {Dobler}, {Lyra},
  {Rheinhardt}, {Bingert}, {Haugen}, {Mee}, {Gent}, {Babkovskaia}, {Yang},
  {Heinemann}, {Dintrans}, {Mitra}, {Candelaresi}, {Warnecke},
  {K{\"a}pyl{\"a}}, {Schreiber}, {Chatterjee}, {K{\"a}pyl{\"a}}, {Li},
  {Kr{\"u}ger}, {Aarnes}, {Sarson}, {Oishi}, {Schober}, {Plasson}, {Sandin},
  {Karchniwy}, {Rodrigues}, {Hubbard}, {Guerrero}, {Snodin}, {Losada},
  {Pekkil{\"a}}, \& {Qian}}]{2021JOSS....6.2807P}
{Pencil Code Collaboration}, {Brandenburg}, A., {Johansen}, A., {et~al.} 2021,
  The Journal of Open Source Software, 6, 2807

\bibitem[{{Porter} \& {Woodward}(2000)}]{2000ApJS..127..159P}
{Porter}, D.~H. \& {Woodward}, P.~R. 2000, \apjs, 127, 159

\bibitem[{{Rempel}(2004)}]{2004ApJ...607.1046R}
{Rempel}, M. 2004, \apj, 607, 1046

\bibitem[{{Roxburgh} \& {Simmons}(1993)}]{1993A&A...277...93R}
{Roxburgh}, L.~W. \& {Simmons}, J. 1993, \aap, 277, 93

\bibitem[{{Scheel} \& {Schumacher}(2016)}]{2016JFM...802..147S}
{Scheel}, J.~D. \& {Schumacher}, J. 2016, Journal of Fluid Mechanics, 802, 147

\bibitem[{{Schumacher} \& {Sreenivasan}(2020)}]{2020RvMP...92d1001S}
{Schumacher}, J. \& {Sreenivasan}, K.~R. 2020, Reviews of Modern Physics, 92,
  041001

\bibitem[{{Singh} \& {Chan}(1993)}]{1993A&A...279..107S}
{Singh}, H.~P. \& {Chan}, K.~L. 1993, \aap, 279, 107

\bibitem[{{Spiegel}(1962)}]{1962JGR....67.3063S}
{Spiegel}, E.~A. 1962, \jgr, 67, 3063

\bibitem[{{Spruit}(1997)}]{Sp97}
{Spruit}, H. 1997, \memsai, 68, 397

\bibitem[{{Tremblay} {et~al.}(2015){Tremblay}, {Ludwig}, {Freytag}, {Fontaine},
  {Steffen}, \& {Brassard}}]{2015ApJ...799..142T}
{Tremblay}, P.-E., {Ludwig}, H.-G., {Freytag}, B., {et~al.} 2015, \apj, 799,
  142

\bibitem[{{Viviani} \& {K{\"a}pyl{\"a}}(2021)}]{2021A&A...645A.141V}
{Viviani}, M. \& {K{\"a}pyl{\"a}}, M.~J. 2021, \aap, 645, A141

\bibitem[{{Weiss} {et~al.}(2004){Weiss}, {Hillebrandt}, {Thomas}, \&
  {Ritter}}]{WHTR04}
{Weiss}, A., {Hillebrandt}, W., {Thomas}, H.-C., \& {Ritter}, H. 2004, {Cox and
  Giuli's Principles of Stellar Structure} (Cambridge, UK: Cambridge Scientific
  Publishers Ltd)

\bibitem[{{Yelles Chaouche} {et~al.}(2020){Yelles Chaouche}, {Cameron},
  {Solanki}, {Riethm{\"u}ller}, {Anusha}, {Witzke}, {Shapiro}, {Barthol},
  {Gandorfer}, {Gizon}, {Hirzberger}, {van Noort}, {Blanco Rodr{\'\i}guez},
  {Del Toro Iniesta}, {Orozco Su{\'a}rez}, {Schmidt}, {Mart{\'\i}nez Pillet},
  \& {Kn{\"o}lker}}]{2020A&A...644A..44Y}
{Yelles Chaouche}, L., {Cameron}, R.~H., {Solanki}, S.~K., {et~al.} 2020, \aap,
  644, A44

\end{thebibliography}
